\documentclass[reprint,superscriptaddress,amsmath,amssymb,prb,nofootinbib,floatfix]{revtex4-2}
\usepackage{graphicx}
\usepackage{dcolumn}
\usepackage{bm}
\usepackage[unicode,colorlinks,urlcolor=blue,linkcolor=blue,pagecolor=blue,citecolor=blue]{hyperref}
\usepackage[english]{babel}
\newcommand\JIHT{Joint Institute for High Temperatures, Izhorskaya 13 Bldg 2, Moscow 125412, Russia}
\newcommand\MIPT{Moscow Institute of Physics and Technology, Institutskiy Pereulok 9, Dolgoprudny, Moscow Region, 141701, Russia}

\begin{document}

%\title{Thermophysical properties of sodium from melting to the critical point: Dependence on the exchange-correlation functional}
\title{A wide-range temperature-dependent deep potential for sodium with near-experimental accuracy from melting to the critical point}
\author{G.\,S.~Demyanov}
\affiliation{\JIHT}
%\affiliation{\MIPT}
\author{D.\,V.~Minakov}
\affiliation{\JIHT}
%\affiliation{\MIPT}
\author{V.\,B.~Fokin}
\affiliation{\JIHT}
%\affiliation{\MIPT}
\author{P.\,R.~Levashov}
\affiliation{\JIHT}
\affiliation{\MIPT}

\date{\today}

\begin{abstract}
	We construct temperature-dependent deep potentials for sodium using finite-temperature DFT data obtained with the PBE, AM05, and $r^2$SCAN exchange--correlation functionals and investigate the thermophysical properties of sodium from room temperature to the critical region. The predicted critical parameters show a dependence on the exchange--correlation functional. The model based on the $r^2$SCAN functional gives a critical temperature $T_c=2.508(8)$~kK, density $\rho_c=0.203(4)$~g/cm$^{3}$, and pressure $P_c=0.249(6)$~kbar, in close agreement with the recommended values. This model is then used to reconstruct the normal-pressure and critical isobars and to calculate the enthalpy, heat capacities, thermal expansion coefficient, bulk moduli, Gr\"uneisen parameter, and speed of sound. The calculated normal-pressure bcc density differs by only 0.06\% from the experimental value. Direct solid--liquid coexistence simulations give a melting temperature of $346(2)$~K, 25~K below the recommended value, consistent with the sensitivity expected from meV/atom free-energy errors. Liquid--vapor coexistence simulations reproduce the binodal and yield a surface tension that approaches zero near the critical point. The calculated self-diffusion coefficient and shear viscosity extend the available transport-property data into the expanded-liquid and near-critical regions, where direct experimental information is sparse. We also demonstrate the thermodynamic consistency of the results by comparing the speed of sound obtained from direct acoustic simulations with that calculated from the equation of state.
	
\end{abstract}

\maketitle

\section{Introduction}

Liquid sodium has long been used as a coolant and heat-transfer medium, particularly in fast-reactor systems \cite{Mathews1993,Sakamoto2013}. Accurate thermophysical data over broad ranges of temperature and pressure are therefore required for the construction of equations of state and for predictive thermal-hydraulic modeling \cite{Li2017,Liu2020}. Near atmospheric pressure, extensive experimental data and recommended correlations are available for the density, enthalpy, heat capacity, speed of sound, viscosity, and surface tension of liquid sodium~\cite{Sobolev2011GENIVCoolants,FinkLeibowitz1995SodiumProperties,Gurvich1978ThermodynamicProperties,narayana2024alkali}. These recommendations draw on calorimetric measurements of the heat capacity and heat of fusion~\cite{Ginnings1950SodiumHeatCapacity}, liquid-enthalpy measurements up to 1505~K~\cite{Fredrickson1974SodiumEnthalpy}, acoustic measurements at elevated pressures~\cite{ShawCaldwell1985SoundAlkali}, and surface-tension measurements~\cite{Poindexter1929SodiumSurfaceTension,Kiriyanenko1965SodiumSurfaceTension}. At higher temperatures, however, particularly in the expanded-liquid and near-critical regimes, direct experimental constraints become increasingly sparse.

Finite-temperature density-functional theory (DFT) and quantum molecular dynamics (QMD) allow one to calculate the thermodynamic properties of liquid metals with electronic excitation taken into account~\cite{Mermin:PR:1965,Kresse:PRB:1993,Kresse:PRB:1996}. Alongside first-principles approaches, classical embedded-atom models have also been used to calculate liquid--vapor coexistence, critical parameters, self-diffusion, and viscosity of sodium~\cite{Metya2012SodiumVLE}.

In our previous papers, QMD calculations were applied to liquid metals near melting and in the critical region~\cite{Minakov:AIPADV:2018,Minakov:PRB:2021,Paramonov:JAP:2022, Minakov:PRB:2018, Minakov:PRB:2022, yl25-2qn3}. However, direct QMD simulations are limited by the accessible system size and trajectory length, which complicates calculations of long-wavelength acoustic modes, transport coefficients from Green--Kubo relations, and two-phase systems. Moreover, meta exchange--correlation functionals, such as the strongly constrained and appropriately normed (SCAN) meta-generalized gradient approximation (meta-GGA), can provide a more accurate description of thermophysical properties than conventional GGA functionals, but at a substantially higher computational cost. For metallic systems, the increased cost per electronic-structure step together with more demanding self-consistent convergence can increase the overall computational expense by up to an order of magnitude, making direct QMD evaluation of thermophysical properties with sufficiently large cells and long trajectories impractical.

First-principles studies of  sodium have addressed several distinct regions of the phase diagram and a variety of liquid-state properties. Qian \textit{et al.} used DFT-based molecular dynamics to study self-diffusion in normal liquid Na and to determine its activation behavior~\cite{Qian1990NaAIMD}. Bickham \textit{et al.} extended \textit{ab initio} molecular dynamics to strongly expanded liquid states, where changes in local coordination and a pronounced decrease in the dc conductivity were found upon expansion~\cite{Bickham1998ExpandedSodium}. First-principles simulations of the liquid--vapor interface subsequently revealed pronounced atomic layering near the triple point~\cite{Gonzalez2004SodiumSurface}. At high pressure, direct DFT molecular dynamics and DFT-trained neural-network potentials were used to investigate the anomalous melting curve and the structural, thermodynamic, and dynamical properties of dense liquid Na~\cite{Hernandez2007SodiumMelting,Koci2008SodiumMelting, 	Eshet2010SodiumNN,Eshet2012SodiumPhaseDiagram, Irie2021SodiumMelting}, while first-principles Kubo--Greenwood calculations provided the electrical and thermal conductivities of the normal liquid at 400~K~\cite{Pozzo2011SodiumConductivity}. \textit{Ab initio} Gibbs-ensemble Monte Carlo simulations mapped liquid--vapor coexistence between 1.2 and 2.3~kK and yielded $T_c=2.338(108)$~kK and $\rho_c=0.24(3)$~g/cm$^3$~\cite{Li2021SodiumAIGEMC}. More recently, direct DFT molecular dynamics has also been used to calculate self-diffusion and shear viscosity in compressed liquid Na~\cite{Liu2025SodiumMelting}. 

The critical parameters of sodium have been estimated by a number of theoretical and semiempirical methods. The available values differ appreciably, especially for the critical density and pressure~\cite{VazquezAlvarezJapas2017SodiumCoexistence,AlvarezJapas2019Sodium,Usov2023SodiumOverheating}. For example, an \textit{ab initio} Gibbs-ensemble Monte Carlo calculation gives $T_c=2.3(1)\times10^3$~K and $\rho_c=0.24(3)$~g/cm$^{3}$~\cite{Li2021SodiumAIGEMC}, while an analysis of liquid--vapor coexistence data gives $\rho_c=0.18(1)$~g/cm$^{3}$~\cite{VazquezAlvarezJapas2017SodiumCoexistence}. Enthalpies, heat capacities, and thermodynamic susceptibilities of the coexisting phases are reconstructed in Ref.~\cite{AlvarezJapas2019Sodium}, and high-temperature extrapolations of the sodium equation of state (EOS) are considered in Ref.~\cite{Usov2023SodiumOverheating}.

These studies, however, generally focus on particular thermodynamic regimes or observables, and a single first-principles-based description combining the equation of state and its derivatives with phase boundaries, interfacial, acoustic, and transport properties from the normal liquid to the near-critical region has remained unavailable.

Machine-learned interatomic potentials provide a route to extending QMD to substantially larger length and time scales while retaining near-DFT accuracy. A particularly relevant example is the transferable machine-learned interatomic potential developed by Kumar \textit{et al.} for aluminum, where a single potential was applied from ambient to warm-dense-matter conditions and used to calculate self-diffusion, shear viscosity, ionic thermal conductivity, sound velocity, and structural properties over an exceptionally broad temperature range~\cite{Kumar2023AlWDM}. Neural-network interatomic potentials have likewise been used to obtain transport coefficients in simulations requiring particularly long or computationally demanding trajectories, including path-integral calculations of hydrogen diffusion on Pd surfaces~\cite{Kataoka2024PdDiffusion}.

Deep neural-network interatomic potentials have also enabled large-scale simulations based on more computationally demanding exchange--correlation approximations. In particular, Luo \textit{et al.} combined the Deep Potential framework with the SCAN meta-GGA functional to investigate structural and diffusive behavior of $\delta$-AlOOH over a broad range of mantle pressures and temperatures~\cite{Luo2024AlOOH}. Temperature-dependent machine-learned interatomic potentials have further been developed for warm dense matter and systems with thermally excited electrons~\cite{Zhang2020TDDPMD,BenMahmoud2022HotElectron,Srinivasan2024eMTP}, while Deep Potential molecular dynamics has recently been applied to the liquid--vapor critical region of aluminum~\cite{LongLuo2026AlCritical}. These studies establish complementary precedents for wide-range machine-learned interatomic potentials for metals, meta-GGA-based Deep Potential simulations, transport calculations, and liquid--vapor criticality. Nevertheless, a unified first-principles-based model that connects the normal and expanded liquid, liquid--vapor coexistence, and the critical region while simultaneously reproducing phase, caloric, acoustic, interfacial, and transport properties remains unexplored.

In this work, we use the Deep Potential method and its implementation in DeePMD-kit~\cite{Zhang2018DeepPotential,Wang2018DeePMDKit} to construct temperature-dependent neural-network interatomic potentials for sodium that reproduce a broad range of thermophysical properties with near-experimental accuracy. In a finite-temperature DFT calculation, the forces and virials are determined from the electronic free energy using the Hellmann--Feynman theorem. A peculiarity of our approach is that, in addition to the free-energy potential, we construct a separate model of the electronic entropy, which is used to recover the internal energy.
Thus, we developed PBE, AM05, and $r^2$SCAN deep potentials from finite-temperature DFT data. We also compare the critical parameters obtained with the three exchange--correlation functionals and use the $r^2$SCAN-based model for the remaining calculations.

Our paper is organized as follows. Section~\ref{sec:methods} introduces the computational methodology. Sections~\ref{sec:dft-training-draft} and~\ref{sec:deep-potential-training-draft} describe the finite-temperature DFT reference calculations, configuration selection, and the construction and training of the Deep Potential models. Section~\ref{sec:md} provides the details of the molecular-dynamics simulations and the calculation of thermodynamic properties, while Sections~\ref{sec:critical-method}, \ref{sec:sound-method}, \ref{sec:phase-methods}, and~\ref{sec:transport-method} describe the methods used to determine the critical point, sound velocity, phase boundaries and surface tension, and transport properties, respectively. Section~\ref{sec:results} presents and discusses the calculated properties of sodium. Section~\ref{sec:eos} presents the equation of state and critical-point analysis, Section~\ref{sec:enthalpy-heat-capacity} discusses the enthalpy and heat capacities, and Section~\ref{sec:mechanical-sound} presents the thermal expansion, bulk moduli, Gr\"uneisen parameter, and speed of sound. Phase boundaries and surface tension are discussed in Section~\ref{sec:phase-results}, while the self-diffusion coefficient and shear viscosity are presented in Section~\ref{sec:dynamics}. Finally, Section~\ref{sec:conclusions} summarizes the main results.

\section{Computational methods}\label{sec:methods}

\subsection{DFT reference calculations and configuration selection}
\label{sec:dft-training-draft}

The reference free energies, forces, and virials are calculated with finite-temperature DFT using the Vienna \textit{Ab initio} Simulation Package (VASP)~\cite{Kresse:PRB:1993,Kresse:PRB:1994,Kresse:PRB:1996,Kresse:CMS:1996}. We use the projector augmented-wave (PAW) method with seven valence electrons~\cite{Blochl:PR:1994,Kresse:PRB:1999}. The electronic occupations are described by the Fermi--Dirac distribution, and the electronic and ionic temperatures are equal. Three exchange--correlation functionals are considered: the Perdew-Burke-Ernzerhof (PBE) parameterization~\cite{Perdew:PRL:1996,Perdew:PRL:1997}, the Armiento-Mattsson (AM05) parameterization~\cite{armiento2005functional}, and $r^2$SCAN~\cite{Furness2020r2SCAN,SunRuzsinszkyPerdew2015SCAN}. The influence of the functional on the equilibrium volume and cohesive properties of alkali metals is discussed in Ref.~\cite{Kovacs2019PBESCANAlkali}.
The plane-wave cutoff is 800~eV to ensure the convergence of pressure and energy. The Brillouin zone is sampled by the Baldereschi mean-value point~\cite{Baldereschi:PRB:1973} for the largest low-density cells and by a $2\times2\times2$ mesh for the smaller cells. All bands with occupations greater than $10^{-5}$ are included.

To generate a broad initial pool of atomic configurations, we first construct a preliminary machine-learning potential from an earlier set of PBE-based QMD simulations covering densities of $\rho=0.05$--$0.90$~g/cm$^{3}$ and temperatures of $T=0.3$--$5.0$~kK. These calculations employ a plane-wave cutoff of 500~eV. The ionic time step is 1--1.5~fs, and each trajectory contains at least 3000 production steps after equilibration. The statistical uncertainty of the averaged thermodynamic quantities is monitored and remains below 1\%. Liquid-state simulations are performed with 128 atoms, except at the two lowest densities, $\rho=0.10$ and $0.05$~g/cm$^{3}$, where 54- and 32-atom cells are used, respectively. From these trajectories, 10\,851 configurations are selected to train the preliminary potential. This model is then used to explore configuration space and generate candidate structures for the subsequent high-accuracy DFT calculations described above.

\begin{figure}
	\centering
	\includegraphics[width=\linewidth]{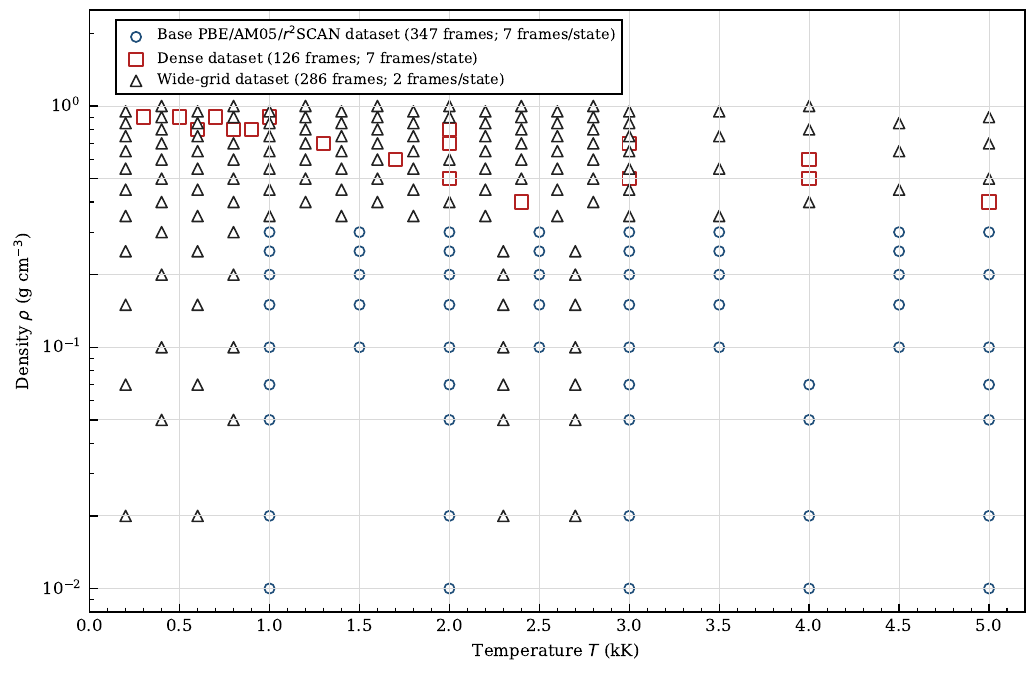}
	\caption{Density--temperature points in the reference datasets. Circles denote the common 347-configuration PBE, AM05, and $r^2$SCAN set. Squares and triangles denote the 126- and 286-configuration $r^2$SCAN extensions, respectively. }
	\label{fig:rhotscandatasets-training-draft}
\end{figure}

The common PBE, AM05, and $r^2$SCAN dataset contains 347 configurations in the ranges $\rho=0.01$--0.30~g/cm$^{3}$ and $T=1$--5~kK and is specifically constructed for a detailed study of the critical region. The simulation cells contain 16--128 atoms. In addition to the MD configurations, isotropically and anisotropically strained cells are included to sample the virial response. The $r^2$SCAN dataset is extended by 126 configurations at $\rho=0.4$--0.9~g/cm$^{3}$ and $T=0.3$--$5.0$~kK and by 286 configurations on a wider grid, $\rho=0.02$--1.00~g/cm$^{3}$ and $T=0.2$--$5.0$~kK. The complete $r^2$SCAN dataset contains 759 configurations (Fig.~\ref{fig:rhotscandatasets-training-draft}).

\subsection{Deep-potential architecture and training}
\label{sec:deep-potential-training-draft}

We construct the DP-PBE, DP-AM05, and DP-$r^2$SCAN models using DeePMD-kit~\cite{Wang2018DeePMDKit}. Each model is trained only on the data obtained with the corresponding exchange--correlation functional. We use the \texttt{se\_e2\_a} descriptor with a cutoff radius of 12~\AA{} and a smoothing radius of 11.2~\AA{}. The embedding network contains layers with 32, 64, and 128 neurons, and the fitting network contains three layers with 240 neurons. The electronic temperature is supplied as a frame parameter. The neighbor-selection parameter is 128 for DP-PBE and DP-AM05 and 384 for DP-$r^2$SCAN. The model parameters are summarized in Table~\ref{tab:dp-training-draft}.

The models are obtained by successive fine-tuning of earlier checkpoints. The final PBE, AM05, and $r^2$SCAN stages contain $5\times10^4$, $10^5$, and $5\times10^4$ optimization steps, respectively. The state points are sampled uniformly. The loss function is
\begin{equation}
	\mathcal{L}
	=
	p_E\mathcal{L}_E
	+
	p_F\mathcal{L}_F
	+
	p_\Xi\mathcal{L}_\Xi,
\end{equation}
where $\Xi$ is the virial. For the final PBE stage, $(p_E,p_F,p_\Xi)$ changes from $(0.1,1000,10)$ to $(1,200,50)$. The corresponding values are $(0.001,1000,10)$ and $(1,200,50)$ for AM05 and $(0.1,2000,1)$ and $(10,50,10)$ for $r^2$SCAN.

The DFT labels correspond to the electronic free energy $F$. To recover the energy without the electronic-entropy term, $E_0$, we train a separate intensive-property model for $s_{\mathrm{el}}=S_{\mathrm{el}}/(Nk_{\mathrm B})$. The descriptor is transferred from DP-$r^2$SCAN, and a new three-layer property network is trained on the 759 configurations for $10^5$ steps. The reference entropy is calculated as $S_{\mathrm{el}}=-(F-E_0)/T$, and
\begin{equation}
	E_0^{\mathrm{pred}}
	=
	F^{\mathrm{pred}}
	+
	Nk_{\mathrm B}Ts_{\mathrm{el}}^{\mathrm{pred}}.
\end{equation}
This model is used only to calculate the entropy contribution. The  forces and virials during the molecular dynamics (MD) simulations are calculated with the free-energy model.

\begin{table}
	\caption{Architecture, training domain, and final fine-tuning parameters of the three deep potentials.}
	\label{tab:dp-training-draft}
	\begin{ruledtabular}
		\begin{tabular}{lccc}
			Model & DP-PBE & DP-AM05 & DP-$r^2$SCAN \\
			\hline
			Descriptor & \multicolumn{3}{c}{\texttt{se\_e2\_a}} \\
			$r_{\rm c}$ (\AA) & 12 & 12 & 12 \\
			$r_{\rm smth}$ (\AA) & 11.2 & 11.2 & 11.2 \\
			Embedding network & \multicolumn{3}{c}{32--64--128} \\
			Fitting network & \multicolumn{3}{c}{240--240--240} \\
			Frame parameters & 1 & 1 & 1 \\
			Neighbor selection & 128 & 128 & 384 \\
			$\rho$ range (g/cm$^{3}$) & 0.01--0.30 & 0.01--0.30 & 0.01--1.00 \\
			$T$ range (kK) & 1.0--5.0 & 1.0--5.0 & 0.2--5.0 \\
			Final steps & $5\times10^4$ & $10^5$ & $5\times10^4$ \\
			%Initial learning rate & $3\times10^{-5}$ & $10^{-4}$ & $10^{-3}$ \\
			%Final learning rate & $10^{-6}$ & $10^{-6}$ & $10^{-6}$ \\
			F RMSE (meV/atom) & 1.80 & 2.15 & 3.53 \\
			Force RMSE (meV/\AA) & 13.3 & 14.3 & 20.0 \\
			Virial RMSE (eV) & 0.286 & 0.282 & 0.679 \\
			P RMSE (kbar) & 0.0149 & 0.0138 & 0.0623 \\
		\end{tabular}
	\end{ruledtabular}
\end{table}

\begin{figure*}[t]
	\centering
	\includegraphics[width=\textwidth]{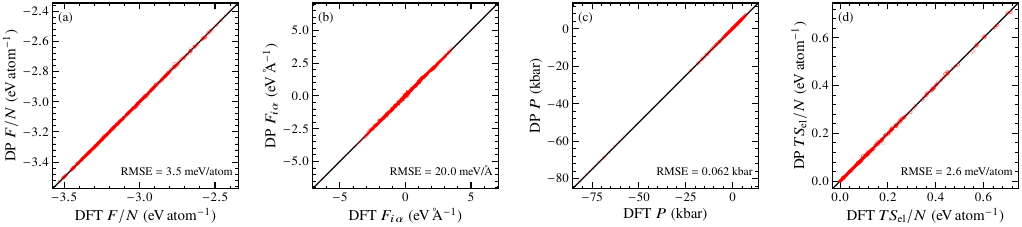}
	\caption{DP-$r^2$SCAN results for the 759-configuration training set: (a) free energy per atom, (b) Cartesian force components, (c) pressure, and (d) electronic entropy contribution per atom. The solid lines correspond to exact agreement.}
	\label{fig:stage9-scan-parity}
\end{figure*}

For DP-$r^2$SCAN, the RMSE values are 3.5~meV/atom for $F/N$, 20.0~meV/\AA{} for the force components, and 0.062~kbar for pressure (Fig.~\ref{fig:stage9-scan-parity}). The RMSE of the entropy model for $TS_{\mathrm{el}}/N$ is 2.6~meV/atom. 
%All 759 configurations are used during optimization; therefore, these values are training errors. 
The relative pressure RMSE does not exceed 3.5\% for $\rho\geq0.05$~g/cm$^{3}$ but reaches 99\% at $\rho=0.01$~g/cm$^{3}$, mainly because of two configurations at 3~kK. The lowest-density states are therefore used only for a qualitative construction of the phase diagram.

\subsection{Molecular-dynamics simulations and thermodynamic properties}\label{sec:md}

Classical MD simulations are performed by LAMMPS~\cite{Thompson2022LAMMPS} using periodic boundary conditions (PBC) with 1~fs time step. Unless stated otherwise, the cubic cell contains 2000 atoms. The EOS is calculated in the $NVT$ ensemble with a Nos\'e--Hoover thermostat and a damping time of 0.1~ps~\cite{Nose:JCP:1984}. Each state is equilibrated for 20~ps and sampled for 50~ps. Configurations for the energy and entropy calculations are stored every 10 steps.

The statistical errors of pressure, energy, and electronic entropy are estimated using block averaging. For pressure, the standard error is calculated as
\begin{equation}
\sigma_P=\left[\frac{1}{n_b(n_b-1)}
\sum_{b=1}^{n_b}\left(\overline{P}_b-\langle P\rangle\right)^2\right]^{1/2},
\label{eq:pressure-block-sem}
\end{equation}
where $n_b=10$, $\overline{P}_b$ is the block average and $\langle P\rangle$ is the mean pressure. Hereafter, the angle brackets are omitted.

The energy provided by DeePMD is the electronic free energy $F$. The total energy is calculated as
\begin{equation}
E=F+T S_{\mathrm{el}}+\frac{3}{2}N k_{\mathrm B}T
=F+N k_{\mathrm B}T s_{\mathrm{el}}+\frac{3}{2}N k_{\mathrm B}T,
\label{eq:md-energy}
\end{equation}
where $s_{\mathrm{el}}=S_{\mathrm{el}}/(N k_{\mathrm B})$. %The free-energy model is used for the trajectories, forces, and pressure, while the entropy model is evaluated for the saved configurations.

The thermodynamic properties are obtained from polynomial approximations of $P(E)|_{\rho}$ and $T(E)|_{\rho}$. At a given pressure $P^\ast$, the isobar is found from the equation $P(E,\rho)=P^\ast$. The derivatives are evaluated analytically from polynomial fits. The enthalpy is
\begin{equation}
H(\rho,T)=E(\rho,T)+P(\rho,T)V,
\qquad
V=\frac{NM}{N_{\mathrm A}\rho},
\label{eq:enthalpy}
\end{equation}
where $M$ is the molar mass of sodium. The heat capacities are calculated as
\begin{equation}
C_V=\left(\frac{\partial E}{\partial T}\right)_V,
\qquad
C_P=\left(\frac{\partial H}{\partial T}\right)_P.
\label{eq:heat-capacities}
\end{equation}
The isobaric expansion coefficient and the Gr\"uneisen parameter are
\begin{equation}
\alpha_P=-\frac{1}{\rho}\left(\frac{\partial\rho}{\partial T}\right)_P
=-\frac{1}{\rho(\partial T/\partial\rho)_P},
\label{eq:alpha-p}
\end{equation}
\begin{equation}
\gamma=V\left(\frac{\partial P}{\partial E}\right)_V.
\label{eq:gruneisen}
\end{equation}
The adiabatic bulk modulus and compressibility are obtained from the thermodynamic speed of sound,
\begin{equation}
K_S=\rho c_{s,\mathrm{th}}^2,
\qquad
\kappa_S=K_S^{-1},
\label{eq:ks}
\end{equation}
and the isothermal quantities are calculated from
\begin{equation}
\kappa_T=\kappa_S+\frac{T\alpha_P^2}{\rho c_P},
\qquad
K_T=\kappa_T^{-1}.
\label{eq:kappa-t}
\end{equation}
Here, $c_P$ denotes the mass-specific isobaric heat capacity.
We also use
\begin{equation}
K_S=K_T\frac{C_P}{C_V}.
\label{eq:ks-ratio}
\end{equation}

Statistical errors are propagated using 250 bootstrap realizations. In each realization, pressure and energy at every state point are varied within their block standard errors, after which the isobars and the derivative quantities are recalculated.

\subsection{Critical-point determination}\label{sec:critical-method}

\begin{figure}
\centering
  \includegraphics[width=0.49\columnwidth]{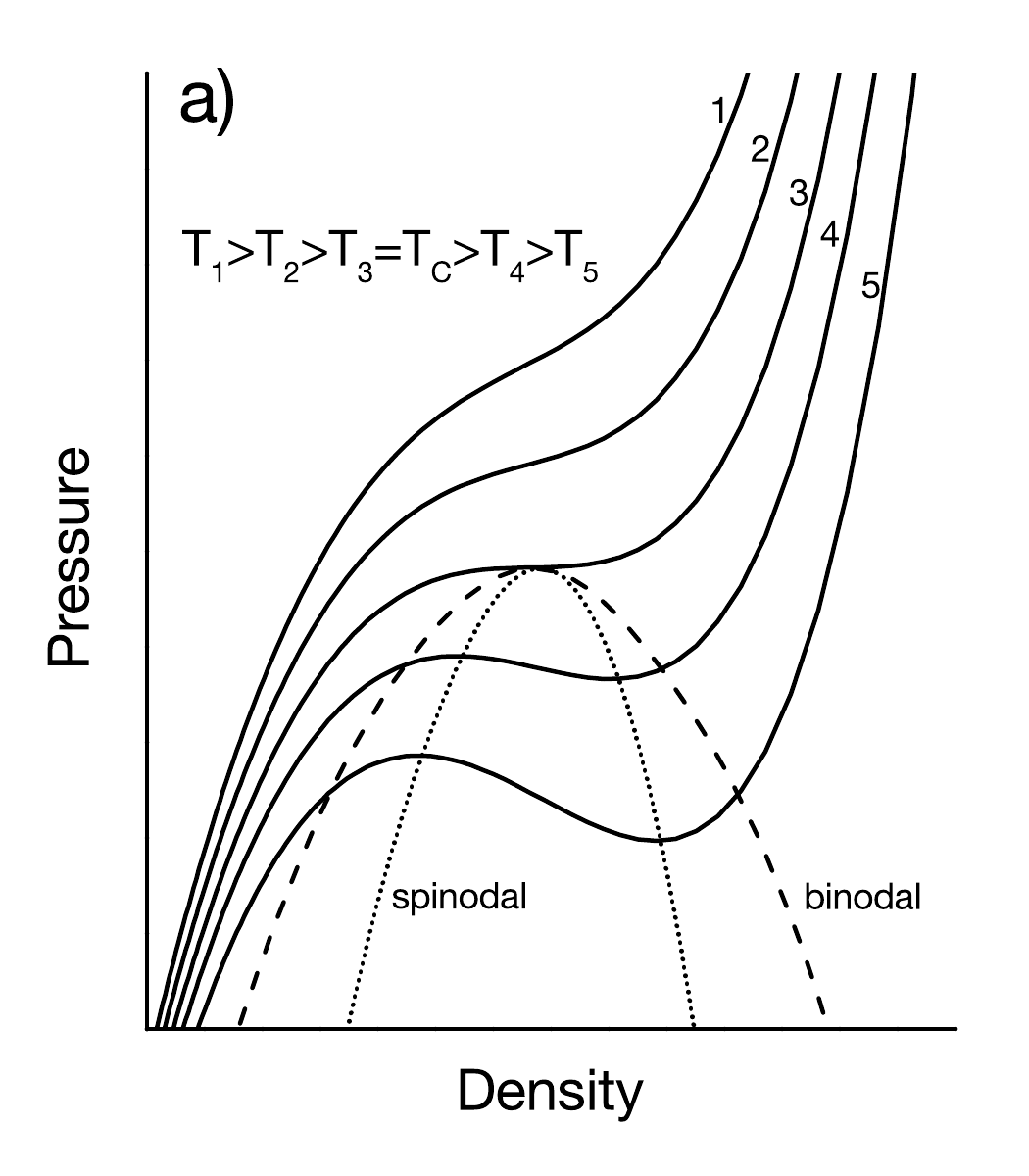}
  \includegraphics[width=0.49\columnwidth]{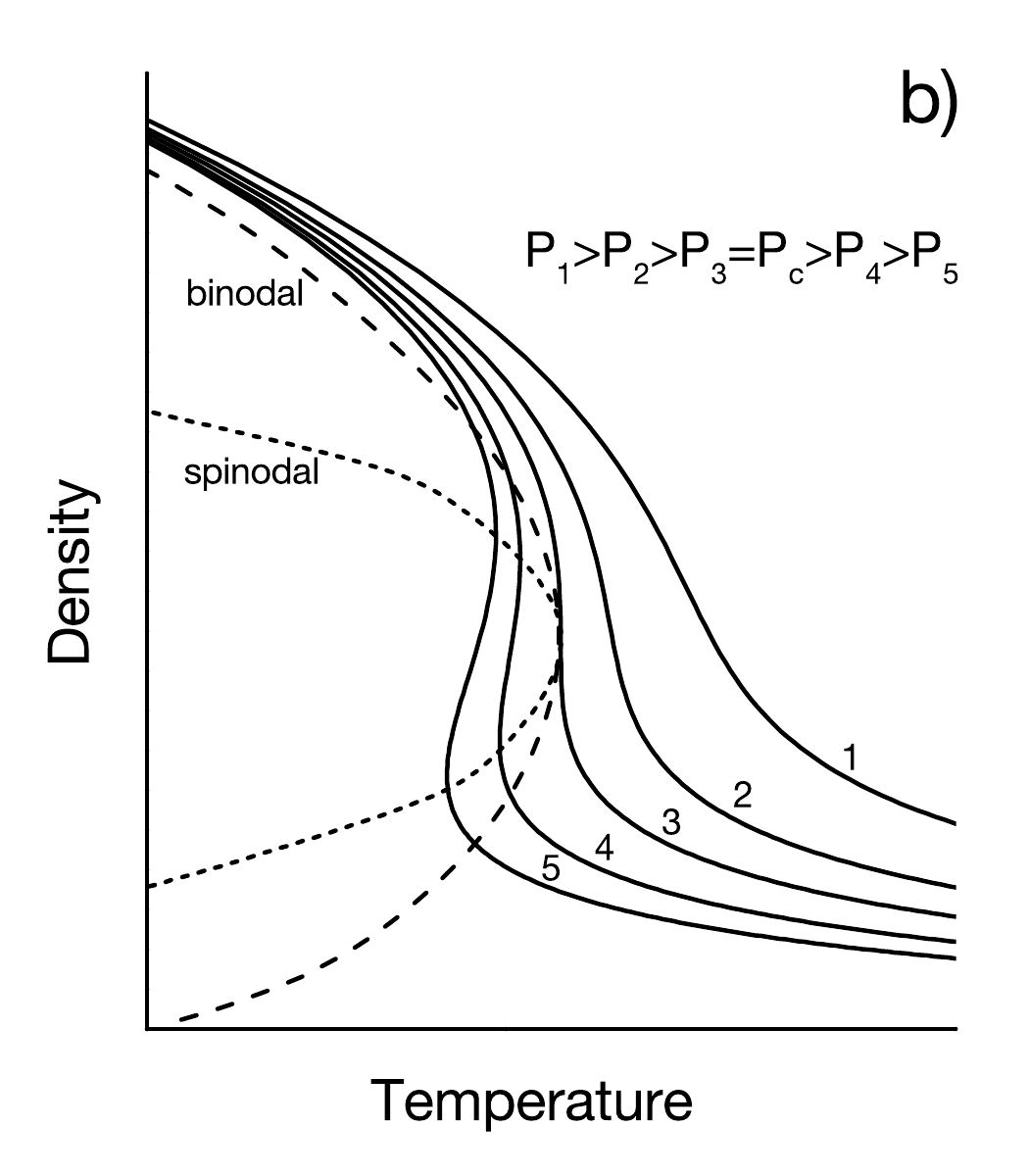}
  \caption{(a) Subcritical and critical isotherms in the $P$--$\rho$ plane. (b) Corresponding isobars in the $\rho$--$T$ plane. The curves are schematic.}
  \label{fig:schem-vdw}
\end{figure}

The critical point is obtained from the pressure surface calculated in the $NVT$ simulations. The construction is based on the approach used in our previous QMD studies~\cite{Minakov:PRB:2021,PhysRevB.110.184204, yl25-2qn3}. First, the pressure along each isochore is approximated as a function of temperature. The resulting fits are then used to reconstruct the pressure isotherms in the $P$--$\rho$ plane. The critical isotherm has a stationary inflection point (Fig.~\ref{fig:schem-vdw}).

For each density $\rho_i$, the pressure is approximated by
\begin{equation}
P_i(T)=a_{i0}+a_{i1}T+a_{i2}T^2.
\end{equation}
The fits are evaluated on a uniform grid of 800 temperatures. At each temperature, the calculated set $\{P_i(T),\rho_i\}$ is approximated by a cubic polynomial,
\begin{equation}
P(\rho;T)=c_0(T)+c_1(T)\rho+c_2(T)\rho^2+c_3(T)\rho^3.
\label{eq:dgtgefsg}
\end{equation}
The inflection density is
\begin{equation}
\rho_*(T)=-\frac{c_2(T)}{3c_3(T)},
\end{equation}
provided that it lies within the fitted density interval. The critical conditions are
\begin{equation}
\left(\frac{\partial P}{\partial\rho}\right)_T=0,
\qquad
\left(\frac{\partial^2P}{\partial\rho^2}\right)_T=0.
\label{eq:critical-conditions}
\end{equation}
We therefore minimize
\begin{equation}
\mathcal{M}(T)=
\left|
\left.
\frac{\partial P}{\partial\rho}
\right|_{\rho=\rho_*(T)}
\right|.
\label{eq:critical-score}
\end{equation}
The minimum gives $T_c$, $\rho_c=\rho_*(T_c)$, and $P_c=P(\rho_c;T_c)$. % At least four isochores are used in each fit, and minima at the boundary of the selected interval are not extrapolated.

The procedure is tested with the reduced van der Waals equation,
\begin{equation}
P_r=\frac{8T_r\rho_r}{3-\rho_r}-3\rho_r^2,
\label{eq:reduced-vdw}
\end{equation}
whose critical coordinates are equal to unity. For $0.90\leq\rho_r\leq1.10$ and $0.98\leq T_r\leq1.05$, the fitted deviations are $T_{r,c}-1=3\times10^{-6}$, $\rho_{r,c}-1=-1.5\times10^{-3}$, and $P_{r,c}-1=6\times10^{-6}$.

For sodium, the critical-point fits use $0.15\leq\rho\leq0.26$~g/cm$^{3}$. The temperature intervals are 2100--2400~K for DP-PBE and DP-AM05, and 2440--2600~K for DP-$r^2$SCAN. Statistical errors are calculated from 500 bootstrap realizations in which the pressure at each state point is varied within its block standard error.

\subsection{Sound-velocity calculation}\label{sec:sound-method}

The adiabatic speed of sound is calculated from the EOS and directly from the propagation of a longitudinal acoustic mode. Along the normal-pressure (1 bar) isobar up to the normal boiling temperature and along the critical isobar, the thermodynamic value is obtained from
\begin{equation}
c_{s,\mathrm{th}}^2=
\left(\frac{\partial P}{\partial\rho}\right)_S
=\frac{K_S}{\rho}
=\frac{\gamma c_P}{\alpha_P},
\label{eq:sound-thermodynamic}
\end{equation}
where $c_P$ is the mass-specific heat capacity. This relation is also used in Refs.~\cite{Lomonosov:2017,Medvedev_2012}.

To verify thermodynamic consistency, we also use a direct method for calculating the speed of sound. For the direct calculation~\cite{10.1063/5.0024150}, the 2000-atom cubic cell is replicated four times along the $x$ direction. The resulting 8000-atom cell is equilibrated for 50~ps in the $NVT$ ensemble. The thermostat is then removed, and the longitudinal perturbation
\begin{equation}
v_x(x,0)=u_0\sin(kx),
\quad
u_0=0.2~\text{\AA/ps},
\quad
k=\frac{2\pi}{L_x}
\label{eq:acoustic-perturbation}
\end{equation}
is applied. The system is subsequently simulated for 150~ps in the microcanonical ensemble.

\begin{figure*}
\centering
\includegraphics[width=0.96\textwidth]{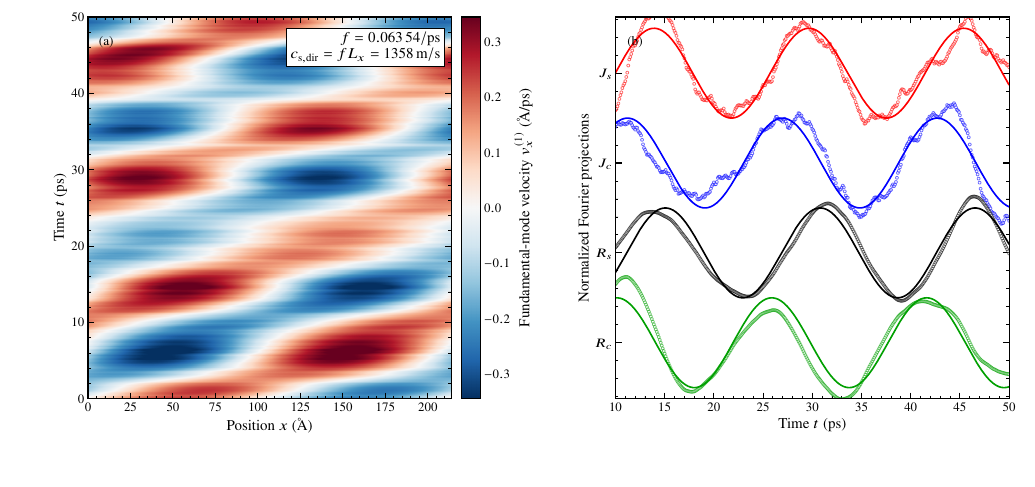}
\caption{Direct speed-of-sound calculation at $\rho=0.5$~g/cm$^{3}$ and $T=2112$~K (critical isobar). (a) Fundamental longitudinal velocity field reconstructed from $J_s(t)$ and $J_c(t)$. (b) Velocity and density projections from Eq.~\eqref{eq:acoustic-projections}. Symbols are MD data, and lines are the joint fits over 10--50~ps.}
\label{fig:direct-acoustic-method}
\end{figure*}

The fundamental mode is determined from the velocity and density projections
\begin{equation}
\begin{aligned}
J_s(t)&=\left\langle v_x\sin(kx)\right\rangle,&
J_c(t)&=\left\langle v_x\cos(kx)\right\rangle,\\
R_s(t)&=\left\langle\sin(kx)\right\rangle,&
R_c(t)&=\left\langle\cos(kx)\right\rangle.
\end{aligned}
\label{eq:acoustic-projections}
\end{equation}
The four projections are fitted simultaneously by damped harmonic functions with a common angular frequency $\omega$ and damping rate. The finite-wave-number phase velocity is
\begin{equation}
c_{s,\mathrm{dir}}=\frac{\omega}{k}=fL_x,
\label{eq:direct-sound}
\end{equation}
where \(f=\omega/(2\pi)\) and \(k=2\pi/L_x\). For the example shown in Fig.~\ref{fig:direct-acoustic-method}, \(f=0.06354/\mathrm{ps}\) and \(L_x=213.80\)~\AA{}, which gives \(c_{s,\mathrm{dir}}=1358\)~m/s.

Only the interval in which the fundamental mode remains coherent is used. The fitting interval is varied between 18 and 30~ps and shifted by 2~ps. The standard deviation of the resulting velocities is taken as the internal uncertainty.

\subsection{Phase-boundary calculations}\label{sec:phase-methods}

%\subsubsection{Melting}\label{sec:melting-method}

A segment of the melting curve near atmospheric pressure is determined from direct solid--liquid coexistence simulations in the $NVT$ ensemble. The fixed-volume periodic cell contains 19800 atoms, with equal solid and liquid parts. The liquid part is prepared at 600~K while the solid part is fixed; the solid is then released, and the whole cell is simulated at the target temperature with a Nos\'e--Hoover thermostat. The first 10~ps are discarded, followed by a 50~ps production run.

Solid-like atoms are identified using the local bcc order with a centrosymmetry parameter below 3.0. From the solid-fraction profile $s(z,t)$, we calculate
\begin{equation}
	L_{\mathrm{s}}(t)=\int s(z,t)\,dz,
	\qquad
	v_{\mathrm{int}}=\frac{1}{2}\frac{dL_{\mathrm{s}}}{dt},
	\label{eq:interface-velocity}
\end{equation}
where the factor $1/2$ accounts for the two interfaces (due to PBC). Positive and negative $v_{\mathrm{int}}$ correspond to freezing and melting, respectively. For each density, $T_{\mathrm m}$ is obtained by linear interpolation to $v_{\mathrm{int}}=0$~\cite{SunAstaHoyt2004InterfaceVelocity}. The corresponding $P_m$ is determined from the interpolated
pressure-tensor component $P_{zz}$ normal to the interface. Statistical uncertainties are propagated by Monte Carlo sampling of the block errors of $v_{\mathrm{int}}$ and $P_{zz}$.
%\subsubsection{Liquid--vapor coexistence and surface tension}\label{sec:coexistence-method}

Liquid--vapor coexistence is simulated in a rectangular periodic cell containing 11\,664 atoms with the mean density 0.19~g/cm$^{3}$. A liquid slab normal to the $z$ axis forms two liquid--vapor interfaces~\cite{Sides1999CapillaryWaves}. The cell volume is fixed.

The calculations are performed at several temperatures between 496 and 2480~K. The low-temperature states are initialized independently. Starting from 2100~K, each state is initialized from the equilibrated configuration at the preceding temperature. Each system is equilibrated for 250~ps and sampled for 500~ps in the $NVT$ ensemble.

The density profile is accumulated in 1~\AA{} bins and recentered using the phase of its first Fourier component. The block-averaged profile is approximated by~\cite{BauerPatel2009WaterVLE}
\begin{equation}
\rho(z)=\rho_v+\frac{\rho_l-\rho_v}{2}
\left[
\tanh\left(\frac{z-z_1}{d_1}\right)
-\tanh\left(\frac{z-z_2}{d_2}\right)
\right],
\label{eq:slab-density-profile}
\end{equation}
where $\rho_l$ and $\rho_v$ are the coexistence densities, $z_1$ and $z_2$ are the interface positions, and $d_1$ and $d_2$ are the interface-width parameters. %The bulk densities are also estimated from the Voronoi volumes.

The surface tension is calculated from the pressure-tensor anisotropy~\cite{KirkwoodBuff1949SurfaceTension},
\begin{equation}
\sigma=\frac{L_z}{2}
\left[P_{zz}-\frac{P_{xx}+P_{yy}}{2}\right].
\label{eq:surface-tension-mechanical}
\end{equation}
%The saturated vapor pressure is taken as
%\begin{equation}
%P_{\mathrm{sat}}(T)=\langle P_{zz}\rangle.
%\label{eq:slab-psat}
%\end{equation}
The pressure-tensor components are stored every 5~fs. The standard error of $\sigma$ is calculated from eight 50-ps blocks
after discarding the first 100~ps of the production interval. Errors of the coexistence densities and interface parameters are obtained from ten profile blocks.

\subsection{Transport properties}\label{sec:transport-method}

The self-diffusion coefficient and shear viscosity are calculated with DP-$r^2$SCAN along the critical isobar from calculations along the isochores. The self-diffusion coefficient is obtained from the long-time slope of the mean-squared displacement,
\begin{equation}
D=\lim_{t\rightarrow\infty}\frac{1}{6t}
\left\langle\frac{1}{N}\sum_{i=1}^{N}
\left|\mathbf{r}_i(t_0+t)-\mathbf{r}_i(t_0)\right|^2
\right\rangle_{t_0}.
\label{eq:self-diffusion}
\end{equation}
The trajectories are unwrapped across periodic boundaries before the mean-squared displacement is calculated.

The shear viscosity is calculated from the Green--Kubo relation,
\begin{equation}
\begin{aligned}
\eta_{\alpha\beta}(t)&=\frac{V}{k_{\mathrm B}T}\int_0^t
\left\langle\delta P_{\alpha\beta}(0)\delta P_{\alpha\beta}(t')\right\rangle\,dt',
\\[-1pt]
&\hspace{3.0em}\alpha\beta\in\{xy,xz,yz\},
\end{aligned}
\label{eq:green-kubo-viscosity}
\end{equation}
where $\delta P_{\alpha\beta}=P_{\alpha\beta}-\langle P_{\alpha\beta}\rangle$. For each state, eight independent $NVT$ simulations containing 2000 atoms are performed with a time step of 0.5~fs. The equilibration and production times are 25 and 200~ps, respectively; the equilibration time is increased to 50~ps in the critical region. The stress tensor is stored every 5~fs.

The three off-diagonal running integrals are averaged as
\begin{equation}
\eta(t)=\frac{1}{3}\left[
\eta_{xy}(t)+\eta_{xz}(t)+\eta_{yz}(t)\right].
\label{eq:viscosity-components}
\end{equation}
The viscosity is determined from the mean value over the 8--20~ps plateau. The integration interval is varied between 6 and 20~ps. 
%For the normal-pressure diffusion coefficients and viscosities, the 95\% confidence intervals are obtained from 5000 bootstrap resamples of the eight replica values. The errors of the critical-isobar diffusion coefficients are estimated by varying the long-time fitting interval.

\section{Results and discussion}\label{sec:results}

\subsection{Equation of state and critical point}\label{sec:eos}\label{sec:critical-results}

The equilibrium density of bcc sodium at 300~K is determined from eight $NVT$ calculations in the interval $\rho=0.965$--0.972~g/cm$^{3}$. In this interval, the pressure is approximated by
\begin{equation}
P(\rho)=A\rho+B.
\end{equation}
Here, $P$ is in kbar and $\rho$ is in g/cm$^{3}$. The coefficients are $A=66.81$~kbar~cm$^{3}$/g and $B=-64.61$~kbar. The normal-pressure density is $\rho_0=0.96709$~g/cm$^{3}$ (lattice parameter $a_0=4.28992$~\AA{}). The density obtained from the experimental lattice parameter $a_0=4.2908$~\AA{} at 298~K is 0.96649~g/cm$^{3}$~\cite{barrett1956alkali,hanfland2002sodium}. Thus, the difference between the calculated and experimental values is 0.06\%.

\begin{figure}
\centering
\includegraphics[width=\columnwidth]{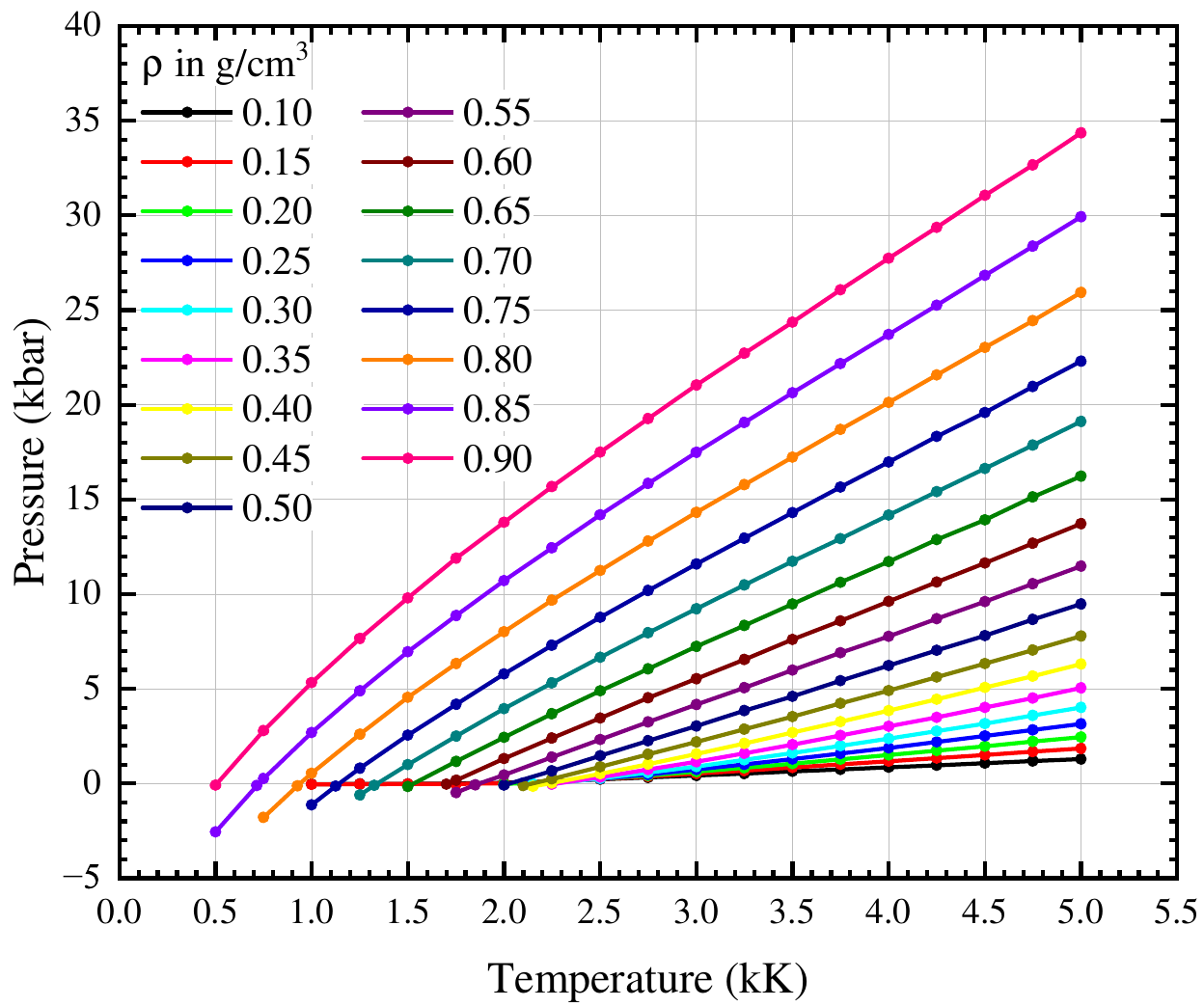}
\caption{Pressure--temperature isochores calculated with DP-$r^2$SCAN. Legend entries give the density in $\mathrm{g/cm^{3}}$. Symbols are $NVT$ averages, and solid lines are the quadratic approximations used to reconstruct the isotherms. Error bars are smaller than the symbol size.}
\label{fig:isochores-pt}
\end{figure}

Using DP-$r^2$SCAN, we calculate pressure and internal energy on a grid of isochores. Figure~\ref{fig:isochores-pt} shows the pressure as a function of temperature. The quadratic approximations of these isochores are used below to reconstruct the isotherms and determine the critical point. The complete EOS data are provided in the Supplemental Material~\cite{supplMat}.

\begin{figure}
\centering
\includegraphics[width=\columnwidth]{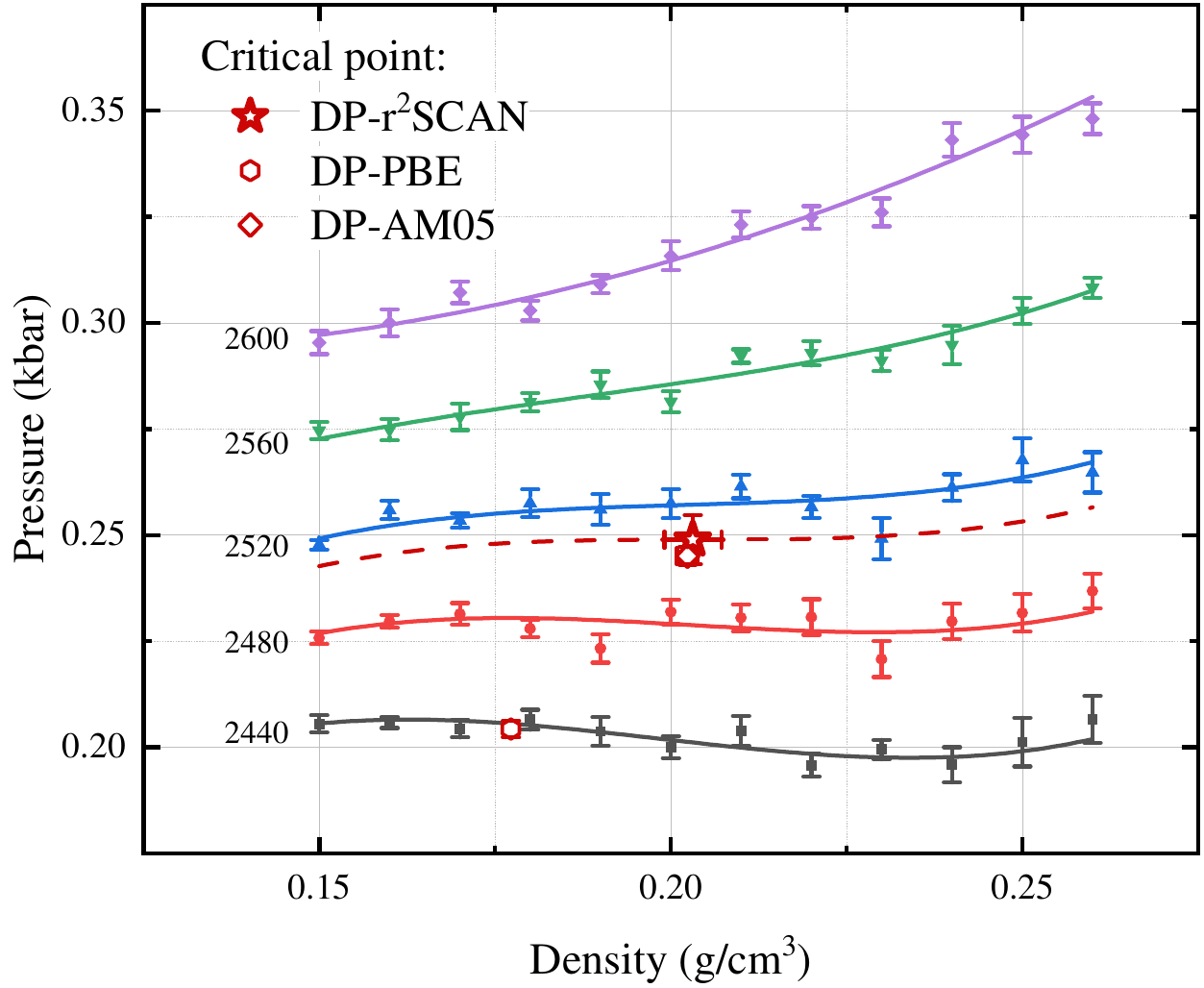}
\caption{Sodium pressure isotherms for DP-$r^2$SCAN near the critical point. Symbols are $NVT$ averages, solid curves are reconstructed from the isochore approximations, and the dashed curve is the critical isotherm. The star, hexagon, and diamond  denote the $r^2$SCAN, PBE, and AM05 critical points, respectively. }
\label{fig:na-critical-results}
\end{figure}

Figure~\ref{fig:na-critical-results} shows the reconstructed $r^2$SCAN isotherms in the vicinity of the critical point. The stationary-inflection method described in Sec.~\ref{sec:critical-method} gives
\begin{equation}
\begin{aligned}
T_c&=2508(8)\ {\rm K},\qquad
\rho_c=0.203(4)\ {\rm g/cm^{3}},\\
P_c&=0.249(6)\ {\rm kbar}.
\end{aligned}
\end{equation}
The corresponding critical compressibility factor is $Z_c=0.135(4)$.

The same procedure gives $T_c=2197(4)$~K, $\rho_c=0.1773(7)$~g/cm$^{3}$, and $P_c=0.204(2)$~kbar for DP-PBE. For DP-AM05, we obtain $T_c=2183(3)$~K, $\rho_c=0.202(1)$~g/cm$^{3}$, and $P_c=0.245(2)$~kbar. Thus, the PBE and AM05 GGA exchange--correlation functionals yield similar critical temperatures, both about 12\% below the Fink--Leibowitz recommendation, although their critical densities differ substantially. In contrast, the meta-GGA $r^2$SCAN functional gives a critical temperature close to the recommended value (Table~\ref{tab:na-critical-points}). For the critical density and pressure, DP-PBE gives the lowest values, whereas DP-AM05 and DP-$r^2$SCAN are very close to each other and show good agreement with the recommended data. The literature estimates in Table~\ref{tab:na-critical-points} vary from 2.3 to 2.80~kK for the critical temperature, from 0.153 to 0.271~g/cm$^{3}$ for the critical density, and from 0.18 to 0.92~kbar for the critical pressure. 

\begin{table}
	\caption{\label{tab:na-critical-points}Critical parameters of sodium from this work and literature. Parentheses denote the reported uncertainties.}
	\setlength{\tabcolsep}{2.5pt}
	\begin{ruledtabular}
		\begin{tabular}{l|ccc}
			\parbox[t]{0.38\columnwidth}{\raggedright Source} &
			$T_c$ (kK) & $\rho_c$ (g/cm$^{3}$) & $P_c$ (kbar)\\
			\hline
			
			\textbf{DP-$r^2$SCAN, this work} &
			\textbf{2.508(8)} &
			\textbf{0.203(4)} &
			\textbf{0.249(6)} \\
			
			\parbox[t]{0.38\columnwidth}{\raggedright DP-PBE, this work} &
			2.197(4) & 0.1773(7) & 0.204(2)\\
			
			\parbox[t]{0.38\columnwidth}{\raggedright DP-AM05, this work} &
			2.183(3) & 0.202(1) & 0.245(2)\\

			\parbox[t]{0.38\columnwidth}{\raggedright Gathers~\cite{Gathers1986DynamicMethods}} &
			2.429 & 0.165 & 0.3\\
			
			\parbox[t]{0.38\columnwidth}{\raggedright Fortov \textit{et al.}~\cite{FortovDreminLeontiev1975CriticalPoint}} &
			2.573 & 0.206 & 0.28\\
			
			\parbox[t]{0.38\columnwidth}{\raggedright Young and Alder~\cite{YoungAlder1971CriticalPointMetals}} &
			2.635 & 0.271 & 0.92\\
			
			\parbox[t]{0.38\columnwidth}{\raggedright Petiot and Seiler~\cite{PetiotSeiler1984SodiumCritical}} &
			2.63(5) & 0.205 & 0.34(4)\\
			
			\parbox[t]{0.38\columnwidth}{\raggedright Hornung~\cite{Hornung1975LiquidMetalCoexistence}} &
			2.573 & 0.221 & 0.35\\
			
			\parbox[t]{0.38\columnwidth}{\raggedright Lomonosov~\cite{Lomonosov2000PhaseDiagrams}} &
			2.473 & 0.24 & 0.466\\
			
			\parbox[t]{0.38\columnwidth}{\raggedright Levashov~\cite{Levashov2000LiquidMetalsDissertation}} &
			2.436 & 0.186 & 0.363\\
			
			\parbox[t]{0.38\columnwidth}{\raggedright Morris~\cite{Morris1964CriticalConstantsMetals}} &
			2.8 & 0.153 & ---\\
			
			\parbox[t]{0.38\columnwidth}{\raggedright Vargaftik~\cite{Vargaftik1975Tables}} &
			2.5 & --- & 0.18\\
			
			\parbox[t]{0.38\columnwidth}{\raggedright Thurnay~\cite{Thurnay1981SodiumProperties}} &
			2.508 & 0.23 & 0.2565\\
			
			\parbox[t]{0.38\columnwidth}{\raggedright Ohse \textit{et al.}~\cite{OhseEtAl1985AlkaliCriticalData}} &
			2.50(2) & 0.211(2) & 0.252(6)\\
			
			\parbox[t]{0.38\columnwidth}{\raggedright Fink and Leibowitz~\cite{FinkLeibowitz1995SodiumProperties}} &
			2.50(1) & 0.22(2) & 0.256(4)\\
			
			\parbox[t]{0.38\columnwidth}{\raggedright Sobolev~\cite{Sobolev2011GENIVCoolants}} &
			2.50(1) & 0.22(2) & 0.256(4)\\
			
			\parbox[t]{0.38\columnwidth}{\raggedright IVTANTHERMO~\cite{Gurvich1978ThermodynamicProperties}} &
			2.50(5) & 0.21(4) & 0.256(2)\\
			
			\parbox[t]{0.38\columnwidth}{\raggedright Vazquez \textit{et al.}~\cite{VazquezAlvarezJapas2017SodiumCoexistence}} &
			--- & 0.18(1) & ---\\
			
			\parbox[t]{0.38\columnwidth}{\raggedright Li \textit{et al.}~\cite{Li2021SodiumAIGEMC}} &
			2.3(1) & 0.24(3) & ---\\
			
			\parbox[t]{0.38\columnwidth}{\raggedright Barnes (estimated)~\cite{Barnes1975SodiumEOS}} & 2.733 & 0.1796 & 0.413\\
			
			\parbox[t]{0.38\columnwidth}{\raggedright Mozgovoi~\textit{et al.}~\cite{Mozgovoi1984CriticalParameters}} &
			2.503(50) & 0.207(30) & 0.256(15)\\

			%\parbox[t]{0.38\columnwidth}{\raggedright Bystrov et al.~\cite{Bystrov1990LiquidMetalCoolants}} & 2.503 & 0.207 & 0.256\\
			
		\end{tabular}
	\end{ruledtabular}
\end{table}

\begin{figure*}
	\centering
	\includegraphics[width=0.8\linewidth]{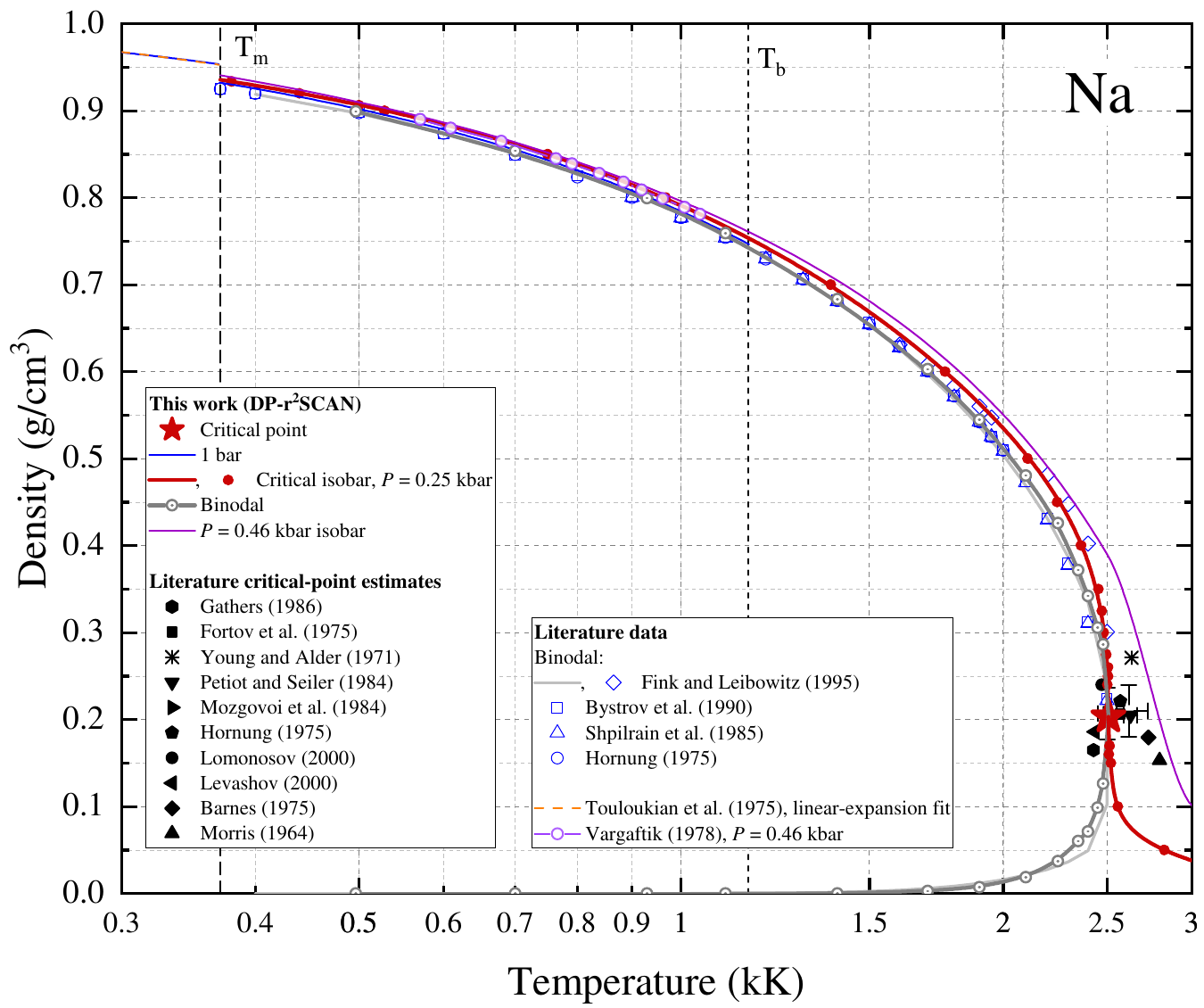}
	\caption{Density--temperature phase diagram of sodium calculated with DP-$r^2$SCAN. The red star denotes the critical point. The red, blue, and violet curves are the $P=0.25$~kbar, normal-pressure, and $P=0.46$~kbar isobars, respectively. Gray circles are the calculated liquid--vapor coexistence densities. The light-gray curve shows the Fink--Leibowitz binodal~\cite{FinkLeibowitz1995SodiumProperties}; saturation-line data from Bystrov~\textit{et  al.}, Shpil'rain~\textit{et  al.}, and Hornung are also shown~\cite{Bystrov1990LiquidMetalCoolants,Shpilrain1985Density,Hornung1975LiquidMetalCoexistence}.
		The Vargaftik data correspond to the $P=470$~kgf/cm$^2$ ($0.46$~kbar) isobar~\cite{Vargaftik1978EOSAlkaliMetals}. 
		The Touloukian curve is obtained from the recommended linear thermal-expansion data for solid sodium~\cite{Touloukian1975ThermalExpansion}.
		Symbols denote selected literature data and critical-point estimates identified in the legend (see Table~\ref{tab:na-critical-points}). Vertical dashed lines mark the experimental normal melting and boiling temperatures.}
	\label{fig:nacritpointphasescan}
\end{figure*}

After the critical point is determined, the liquid branch of the critical isobar is calculated (see Fig.~\ref{fig:nacritpointphasescan}) and approximated by
\begin{equation}
\rho(T)=\rho_c+C_1u+C_2u^2+C_3u^3+C_4u^4+C_5u^5,
\quad P=P_c,
\label{eq:rho-critical-isobar}
\end{equation}
where $u=(T_c-T)^{1/3}$, with both $T$ and $T_c$ expressed in kK, and $\rho$ is in g/cm$^{3}$. The coefficient $C_n$ is in $\mathrm{g/(cm^{3}\,kK^{n/3})}$. The coefficients are $C_1=0.1669$, $C_2=0.9729$, $C_3=-1.731$, $C_4=1.432$, and $C_5=-0.3782$.
The approximation is used in the interval $0.380~\mathrm{kK}\leq T< T_c$.

%Figure~\ref{fig:nacritpointphasescan} shows the density--temperature phase diagram obtained with DP-$r^2$SCAN. The critical point and the normal-pressure and critical isobars discussed above are shown together with the calculated coexistence densities and available literature data. The coexistence calculations are discussed in Sec.~\ref{sec:interface}. 
%The DP-$r^2$SCAN critical point is close to the Fink--Leibowitz estimate and has a lower density than the Lomonosov critical point~\cite{Lomonosov2000PhaseDiagrams,FinkLeibowitz1995SodiumProperties}.

Figure~\ref{fig:nacritpointphasescan} summarizes the density--temperature phase diagram obtained with DP-$r^2$SCAN. The calculated atmospheric-pressure isobar agrees closely with the available literature data from the vicinity of the melting point up to the normal boiling temperature. The $P=0.46$~kbar isobar also follows the data of Vargaftik~\cite{Vargaftik1978EOSAlkaliMetals}.

The calculated coexistence densities reproduce the overall shape of the liquid--vapor binodal and are close to the Fink--Leibowitz~\cite{FinkLeibowitz1995SodiumProperties} correlation over both branches.  Near the critical region, the present DP-$r^2$SCAN critical point lies within the broad scatter of previous estimates and close to the recommended values listed in Table~\ref{tab:na-critical-points}. The coexistence calculations are discussed in Sec.~\ref{sec:phase-results}.

\subsection{Enthalpy and heat capacities}\label{sec:enthalpy-heat-capacity}

\begin{figure}
	\centering
	\includegraphics[width=\columnwidth]{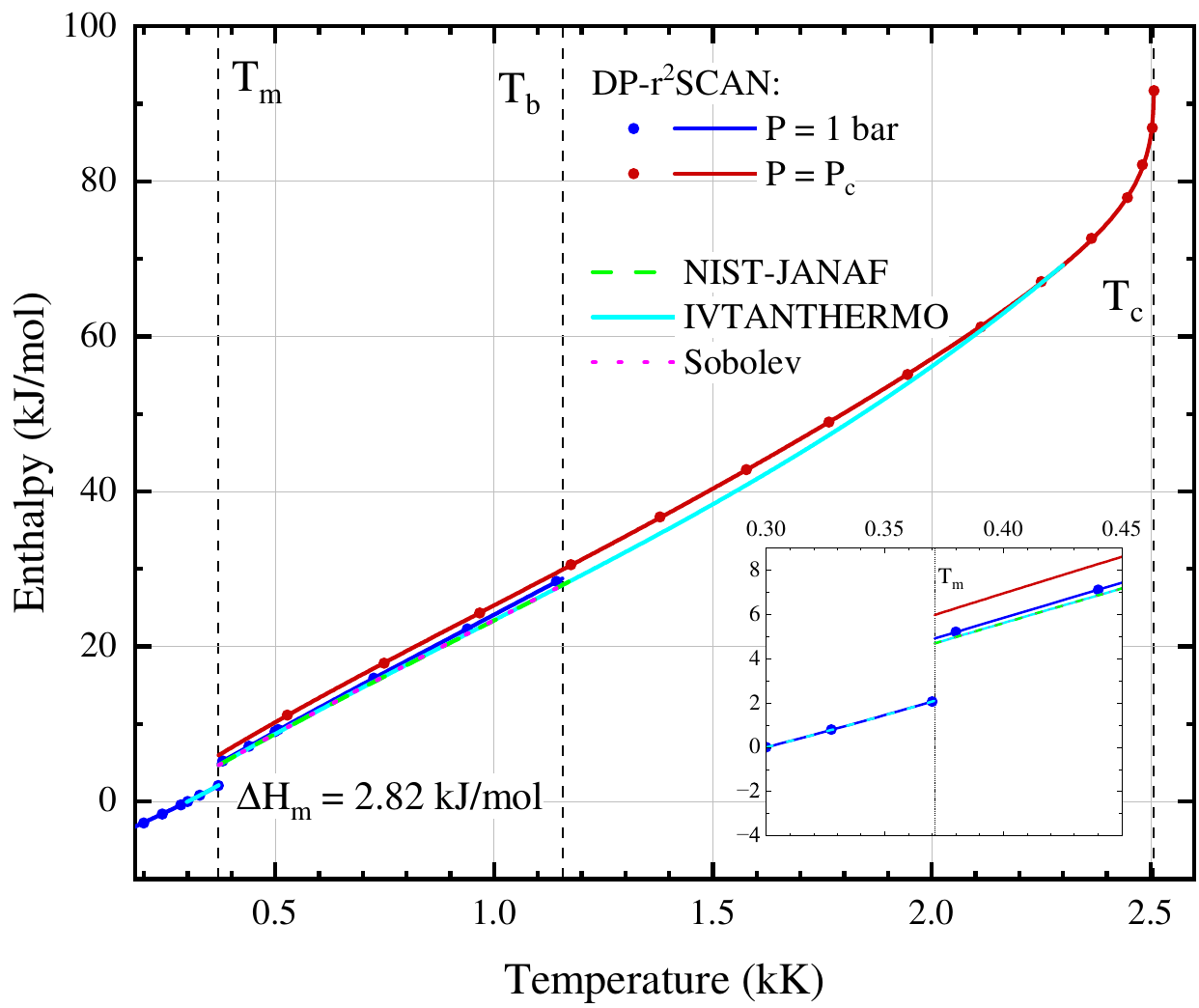}
  \caption{Relative enthalpy of bcc and liquid sodium at normal pressure 	and of liquid sodium along the critical isobar. Symbols denote the 	DP-$r^2$SCAN results. The blue and red curves represent the normal-pressure and 	critical-isobar dependences, respectively. The dashed blue segment 	extends the liquid normal-pressure branch from 380~K to the experimental melting 	temperature $T_{\mathrm m}$. The green dashed, cyan solid, and magenta 	dotted curves show the NIST--JANAF, Gurvich--IVTANTHERMO, and Sobolev 	reference dependences, respectively 	~\cite{Chase1998JANAF,Gurvich1978ThermodynamicProperties, 		Sobolev2011GENIVCoolants}. The inset enlarges the region near 	$T_{\mathrm m}$. Vertical dotted lines mark $T_{\mathrm m}$, the normal 	boiling temperature $T_{\mathrm b}$, and the critical temperature 	$T_{\mathrm c}$.}
	\label{fig:enthalpy-isobars}
\end{figure}

Figure~\ref{fig:enthalpy-isobars} shows the calculated enthalpy of bcc and liquid sodium at atmospheric pressure and of liquid sodium along the critical isobar.
For bcc sodium, the normal-pressure enthalpy in the interval
$0.25~\mathrm{kK}\leq T\leq T_{\mathrm m}^{\mathrm{exp}}$
is approximated by
\begin{equation}
	H_0(T)-H_0(300~\mathrm K)
	=
	D_1\theta_0
	+
	D_2\left[\exp(D_3\theta_0)-1\right],
	\label{eq:h-solid-p0}
\end{equation}
where $\theta_0=T-0.3$, $T$ in kK, and the coefficients are
$D_1=24.9434$~kJ~mol$^{-1}$~kK$^{-1}$,
$D_2=0.34802$~kJ/mol, and
$D_3=9.47319$~kK$^{-1}$.
The corresponding isobaric heat capacity is obtained by differentiating
Eq.~\eqref{eq:h-solid-p0},
\begin{equation}
	C_{0}^{\mathrm{bcc}}(T)
	=
	D_1+D_2D_3\exp(D_3\theta_0).
	\label{eq:cp-solid-p0}
\end{equation}

The liquid ambient-pressure enthalpy is approximated over
$T_{\mathrm m}^{\mathrm{exp}}\leq T\leq T_{\mathrm b}$ by
\begin{equation}
	\begin{aligned}
		H_0(T)-H_0(300~\mathrm K)
		={}&F_0+F_1\theta_0
		+F_2\theta_0^2
		+F_3\theta_0^3\\
		&+F_4\exp(F_5\theta_0).
	\end{aligned}
	\label{eq:h-t-fit-p0}
\end{equation}
Here, $T_{\mathrm b}=1156$~K denotes the normal boiling temperature of sodium.
The coefficients are
$F_0=2.62020$~kJ/mol,
$F_1=32.7395$~kJ~mol$^{-1}$~kK$^{-1}$,
$F_2=-4.53996$~kJ~mol$^{-1}$~kK$^{-2}$,
$F_3=2.30458$~kJ~mol$^{-1}$~kK$^{-3}$,
$F_4=8.517\times10^{-11}$~kJ/mol, and
$F_5=12.4737$~kK$^{-1}$.

Over the same interval, differentiation of Eq.~\eqref{eq:h-t-fit-p0} gives
\begin{equation}
	C_{0}(T)
	=
	F_1+2F_2\theta_0
	+3F_3\theta_0^2
	+F_4F_5\exp(F_5\theta_0).
	\label{eq:cp-liquid-p0}
\end{equation}
Eqs.~\eqref{eq:cp-solid-p0} and~\eqref{eq:cp-liquid-p0} give $C_P$
directly in J~mol$^{-1}$~K$^{-1}$.

The calculated bcc enthalpy agrees with the reference data (see Fig.~\ref{fig:enthalpy-isobars}). In the liquid state, the calculated curve has a slightly larger slope than the experimental and recommended dependences. At the experimental melting temperature, the calculated fusion enthalpy is $\Delta H_{\mathrm m}=2.82$~kJ/mol, which is 9\% above the IVTANTHERMO value $2.598(5)$~kJ/mol~\cite{Gurvich1978ThermodynamicProperties}. The inset in Fig.~\ref{fig:enthalpy-isobars} shows that the difference is mainly caused by the calculated liquid enthalpy at melting, whereas the enthalpy of bcc sodium is in excellent agreement with the literature data.

Along the critical isobar, the temperature dependence of enthalpy in the interval
$0.5~\mathrm{kK}\leq T<T_c$ is described by
\begin{multline}
	H_{P=P_c}(T)-H_{P=P_c}(300~\mathrm K)
	=
	G_0+G_1\theta_{\mathrm c}
	+G_2\theta_{\mathrm c}^{2}
	+G_3\theta_{\mathrm c}^{3}\\
	+G_4\theta_{\mathrm c}^{4}
	+G_5\left(-\frac{\theta_{\mathrm c}}{T_c}\right)^{G_6},
	\label{eq:h-t-critical-fit}
\end{multline}
where $\theta_{\mathrm c}=T-T_c$, $T$ and $T_c = 2.50849$ are  in kK.
The coefficients are
$G_0=99.8105$~kJ/mol,
$G_1=19.1072$~kJ~mol$^{-1}$~kK$^{-1}$,
$G_2=-8.37354$~kJ~mol$^{-1}$~kK$^{-2}$,
$G_3=-5.67561$~kJ~mol$^{-1}$~kK$^{-3}$,
$G_4=-1.35200$~kJ~mol$^{-1}$~kK$^{-4}$,
$G_5=-43.2970$~kJ/mol, and
$G_6=0.2$.
The polynomial contribution describes the regular temperature dependence,
whereas the last term accounts for the rapid increase near the critical
point. 

Differentiation of Eq.~\eqref{eq:h-t-critical-fit} gives the isobaric
heat capacity,
\begin{multline}
	C_{P=P_c}(T)=
	G_1+2G_2\theta_{\mathrm c}
	+3G_3\theta_{\mathrm c}^{2}\\
	+4G_4\theta_{\mathrm c}^{3}
	-\frac{G_5G_6}{T_c}
	\left(-\frac{\theta_{\mathrm c}}{T_c}\right)^{G_6-1}.
	\label{eq:cp-critical-fit}
\end{multline}
Again, 
Eq.~\eqref{eq:cp-critical-fit} gives $C_P$ directly in
J~mol$^{-1}$~K$^{-1}$. 
Near the critical point,
$C_P\propto(-\theta_{\mathrm c}/T_c)^{-0.8}$, whereas enthalpy remains finite
and approaches its critical value as
$H_c-H(T)\propto(-\theta_{\mathrm c}/T_c)^{0.2}$.
This approximation is used below together with the density and Gr\"uneisen-parameter dependences to calculate the speed of sound.

\begin{figure}
	\centering
	\includegraphics[width=\columnwidth]{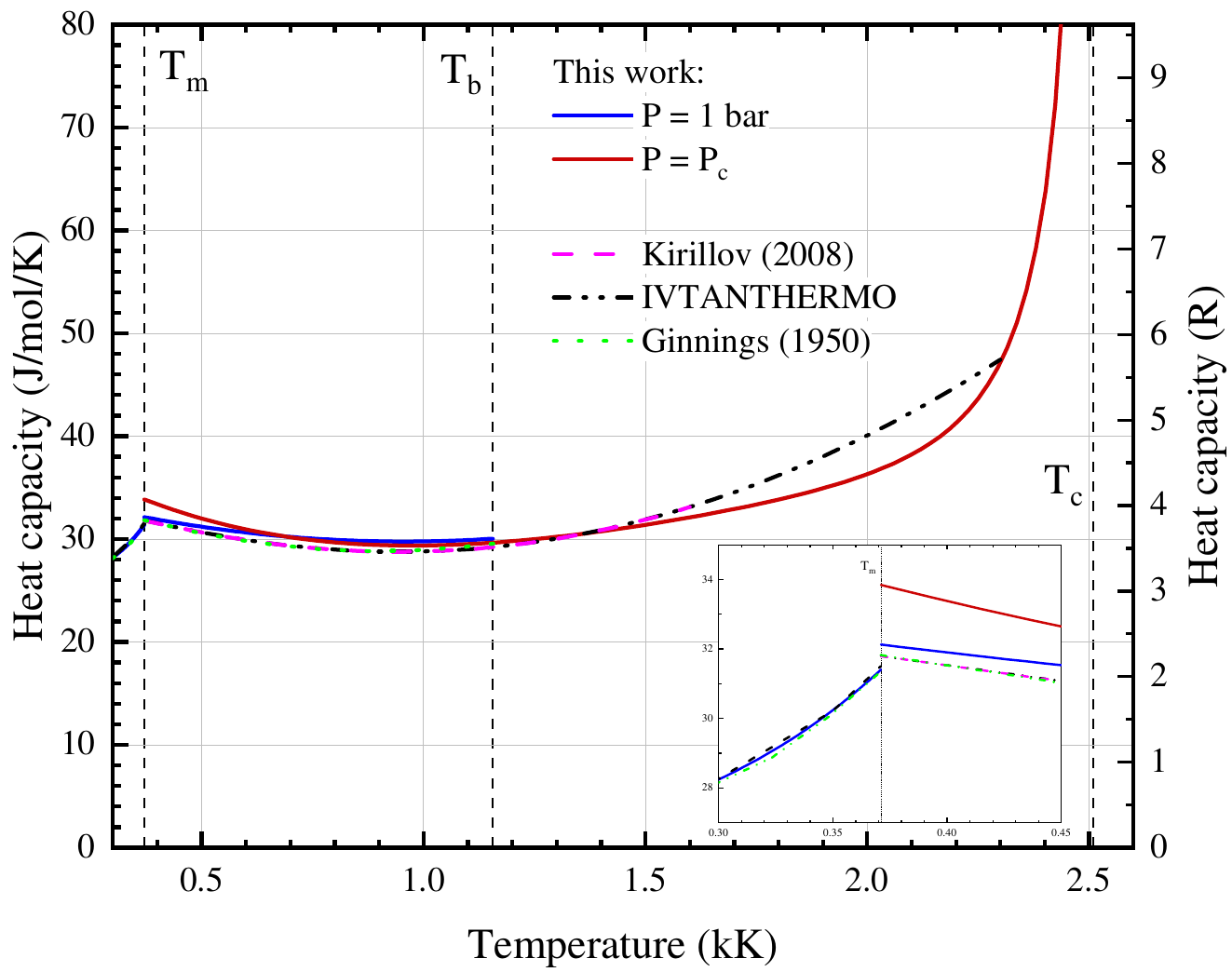}
  \caption{Isobaric heat capacity along the atmospheric-pressure and critical isobars. The blue and red curves are the present DP-$r^2$SCAN results along the atmospheric-pressure and critical isobars, respectively. The magenta dashed, black dash-dotted, and green dotted curves correspond to the Kirillov (2008) recommendation as compiled by Sobolev, Gurvich--IVTANTHERMO, and Ginnings et al. (1950), respectively ~\cite{Kirillov2008ThermophysicalProperties, Sobolev2011GENIVCoolants,Gurvich1978ThermodynamicProperties, Ginnings1950SodiumHeatCapacity}. The inset enlarges the region near the normal melting temperature $T_m$ and shows the discontinuity in $C_P$ between the solid and liquid branches. Vertical dotted lines in the main panel mark $T_m$, the normal boiling temperature $T_b$, and the critical temperature $T_c$.}
	\label{fig:cp-isobars}
\end{figure}

In Fig.~\ref{fig:enthalpy-isobars}, the enthalpy results are compared with the Sobolev, NIST--JANAF, and Gurvich--IVTANTHERMO correlations ~\cite{Sobolev2011GENIVCoolants,Chase1998JANAF, Gurvich1978ThermodynamicProperties}. In Fig.~\ref{fig:cp-isobars}, the heat-capacity results are compared with the Kirillov (2008), Gurvich--IVTANTHERMO, and Ginnings dependences~\cite{Kirillov2008ThermophysicalProperties, Gurvich1978ThermodynamicProperties,Ginnings1950SodiumHeatCapacity}.
Up to approximately 1.15~kK, the calculated liquid enthalpy increases slightly faster than the reference curves. At normal pressure, the calculated $C_P$ agrees with the recommended dependences. Along the critical isobar, $C_P$ increases rapidly as $T_c$ is approached.

The isochoric heat capacity is calculated from $E(T)|_\rho$ as
\begin{equation}
	C_V=\left(\frac{\partial E}{\partial T}\right)_V.
\end{equation}
Along the critical isobar, its temperature dependence is approximated in the range
$0.5~\mathrm{kK}\leq T<T_c$ by 
\begin{equation}
	C_{V,P=P_{\mathrm c}}(T)=A_0+A_1\theta_{\mathrm c}+A_2\theta_{\mathrm c}^2+A_3\theta_{\mathrm c}^3+A_4(-\theta_{\mathrm c})^{A_5}.
	\label{eq:cv-critical-fit}
\end{equation}
Here, $\theta_{\mathrm c}=T-T_{\mathrm c}$, with $T$ and $T_{\mathrm c}$ expressed in kK. The coefficients are $A_0=0.321182$~$k_{\mathrm B}$/atom, $A_1=-0.394858$~$k_{\mathrm B}$\,atom$^{-1}$\,kK$^{-1}$, $A_2=-0.137804$~$k_{\mathrm B}$\,atom$^{-1}$\,kK$^{-2}$, $A_3=-0.141090$~$k_{\mathrm B}$\,atom$^{-1}$\,kK$^{-3}$, $A_4=1.66122$~$k_{\mathrm B}$\,atom$^{-1}$\,kK$^{0.11}$, and $A_5=-0.11$. The resulting $C_V$ is expressed in $k_{\mathrm B}$/atom.
The regular cubic contribution reproduces the broad minimum in the liquid region and the low-temperature slope, whereas the last term describes the increase toward the critical point. 

\begin{figure}
	\centering
	\includegraphics[width=\columnwidth]{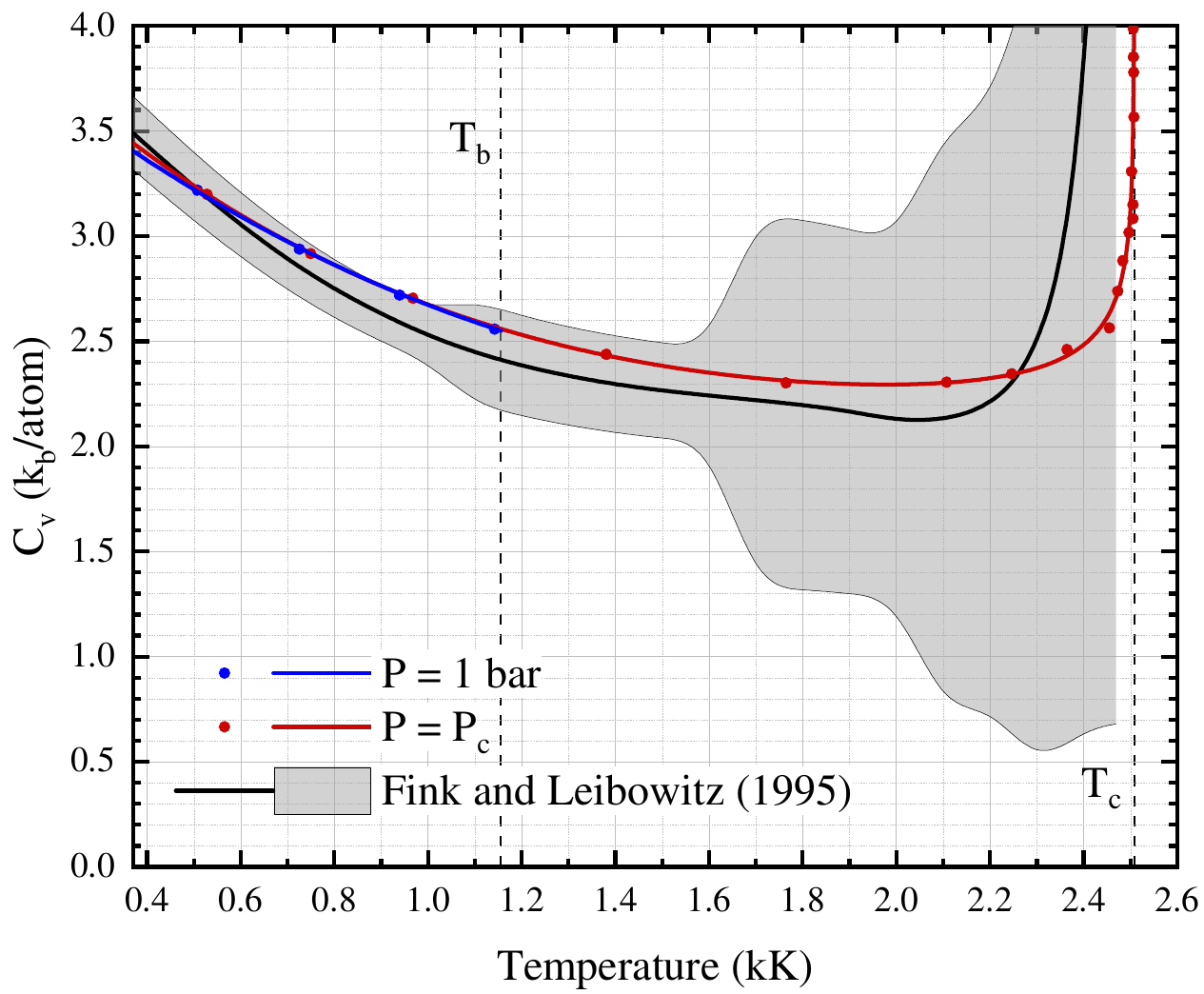}
	\caption{Isochoric heat capacity along the ambient-pressure and critical isobars. Symbols are obtained from the calculated EOS. The critical-isobar curve is described by Eq.~\eqref{eq:cv-critical-fit}. The black curve and gray band show the Fink--Leibowitz recommendation and its uncertainty~\cite{FinkLeibowitz1995SodiumProperties}.}
	\label{fig:cv-isochores}
\end{figure}

Figure~\ref{fig:cv-isochores} shows the temperature dependence of the isochoric heat capacity along the normal-pressure and critical isobars in comparison with the recommended data of Fink and Leibowitz~\cite{FinkLeibowitz1995SodiumProperties}. At low temperatures, the two calculated isobars are nearly indistinguishable and agree with the recommended dependence within its uncertainty up to the boiling temperature. Above $T_{\mathrm b}$, the normal-pressure liquid branch is no longer an equilibrium state; along the critical isobar, $C_V$ exhibits a pronounced rise as $T_c$ is approached. The uncertainty of the recommended data increases rapidly in the expanded-liquid and near-critical regions.

\subsection{Thermal expansion, bulk moduli, Gr\"uneisen parameter, and speed of sound}
\label{sec:mechanical-sound}

\begin{figure}
	\centering
	\includegraphics[width=\linewidth]{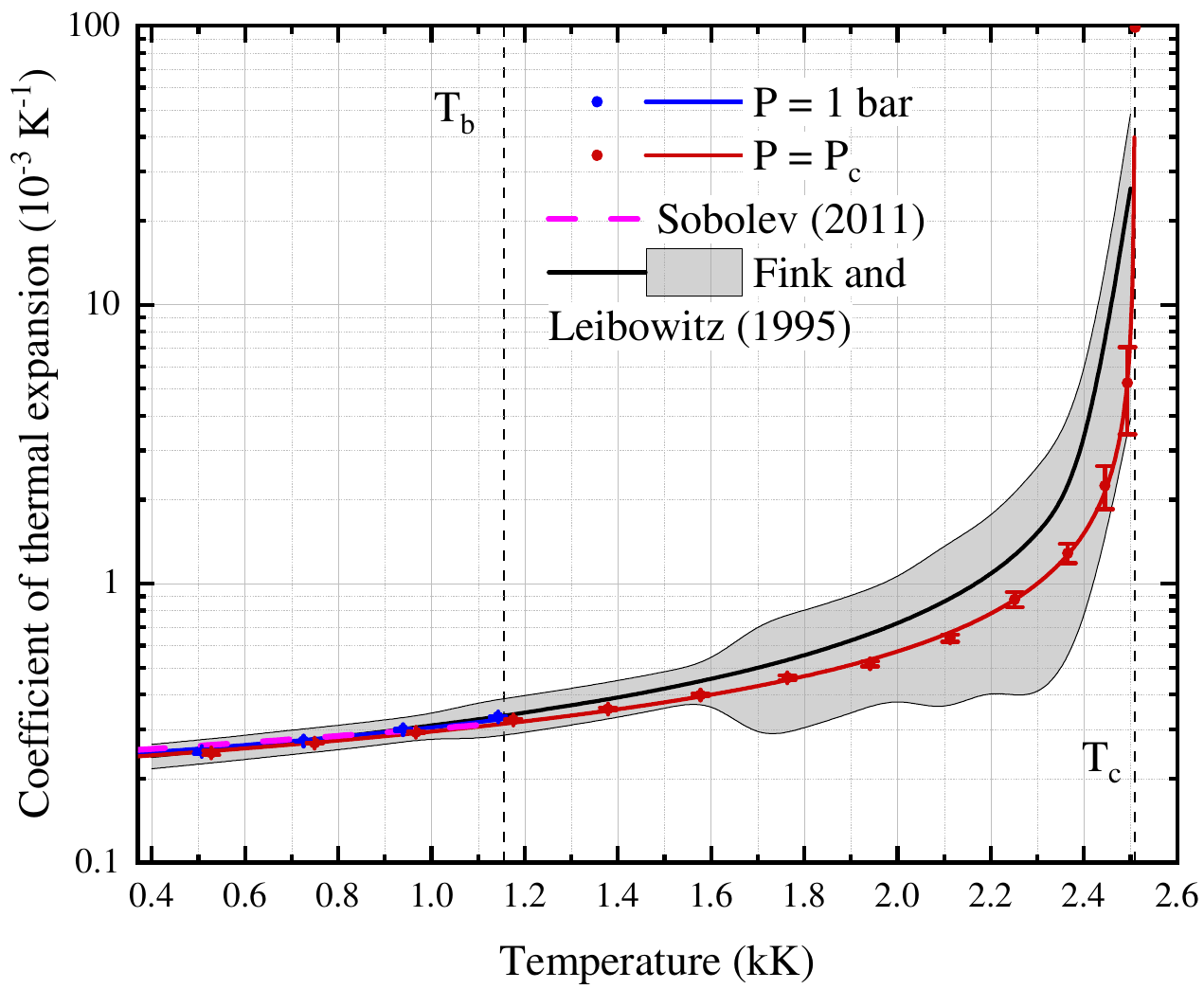}
	\caption{Isobaric thermal expansion coefficient of sodium along the ambient-pressure and critical isobars. Symbols show the DP-$r^2$SCAN results, and solid curves are analytic approximations. The Sobolev normal-pressure correlation~\cite{Sobolev2011GENIVCoolants} and the Fink--Leibowitz saturated-liquid recommendation~\cite{FinkLeibowitz1995SodiumProperties} are shown for comparison. }
	\label{fig:expansion}
\end{figure}

The isobaric thermal expansion coefficient, $\alpha_P$, and the bulk moduli are calculated from the reconstructed EOS along the atmospheric-pressure and critical isobars. Figure~\ref{fig:expansion} shows the temperature dependence of $\alpha_P$. At low temperatures, the two isobars are nearly indistinguishable and agree closely with the normal-pressure Sobolev recommendation~\cite{Sobolev2011GENIVCoolants}. Along the atmospheric-pressure isobar, $\alpha_P$ increases from $2.5\times10^{-4}$~K$^{-1}$ at melting to $3.3\times10^{-4}$~K$^{-1}$ at the normal boiling temperature $T_b$. Along the critical isobar, the increase becomes substantially stronger near the critical region, reaching $5.24\times10^{-3}$~K$^{-1}$ at 2.493~kK. The rapid growth of $\alpha_P$ reflects the increasing sensitivity of the density to temperature as the critical point is approached.

The Fink--Leibowitz dependence~\cite{FinkLeibowitz1995SodiumProperties} for saturated liquid sodium is also shown in Fig.~\ref{fig:expansion}. It agrees with the present results below $T_{\mathrm b}$, where the differences between the ambient-pressure, critical-isobar, and saturation curves are small. At higher temperatures, the uncertainty of the Fink--Leibowitz recommendation increases considerably.

\begin{figure}
	\centering
	\includegraphics[width=\linewidth]{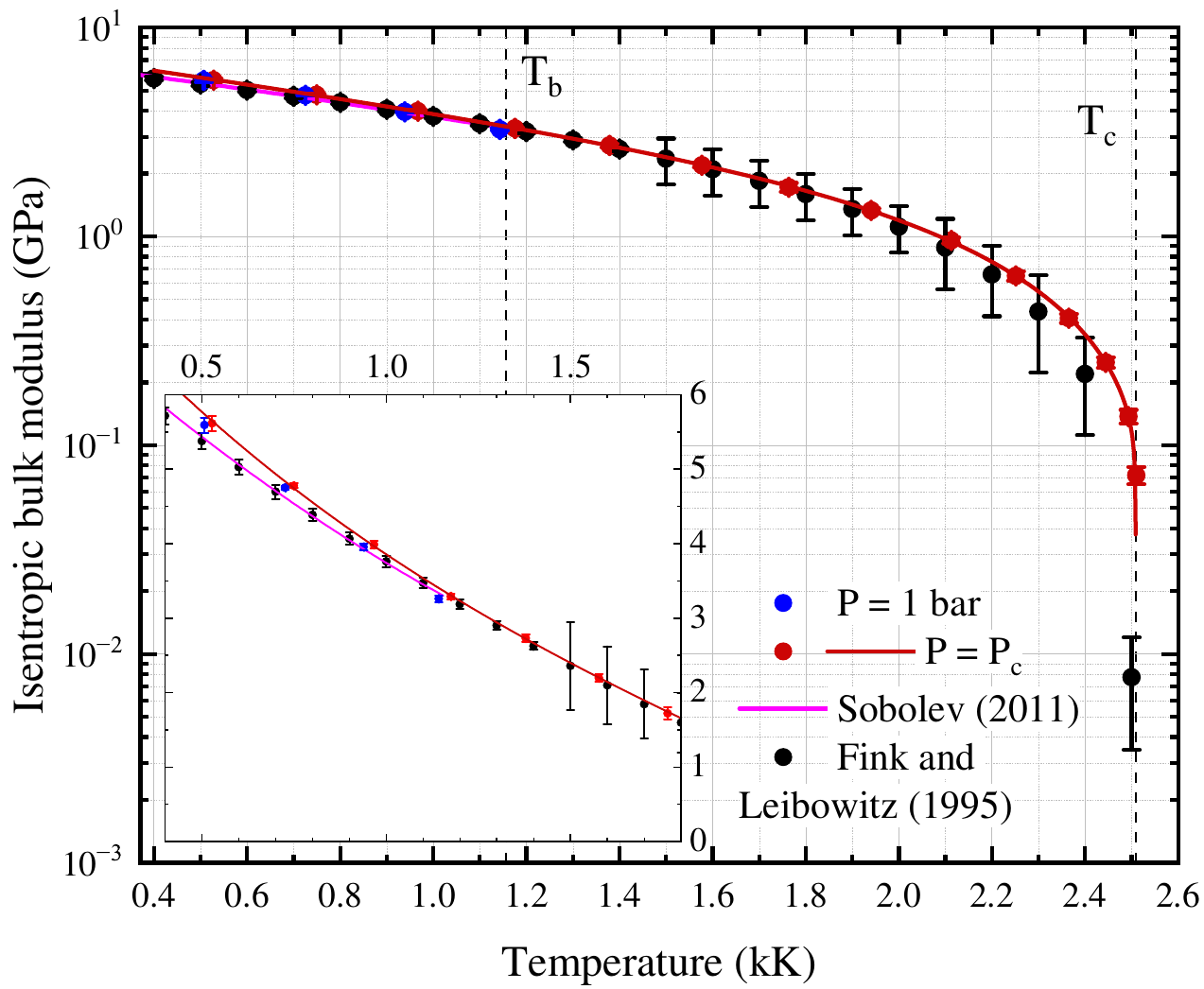}
	\caption{Isentropic bulk modulus $K_S$ of sodium along the ambient-pressure and critical isobars. Symbols show the DP-$r^2$SCAN results, and the solid curves are analytic approximations. Black symbols are values obtained from the Fink--Leibowitz saturated-liquid compressibility~\cite{FinkLeibowitz1995SodiumProperties}, and the magenta curve is the Sobolev normal-pressure dependence~\cite{Sobolev2011GENIVCoolants}.  The inset shows the low-temperature region on a linear ordinate scale.}
	\label{fig:Ks}
\end{figure}

\begin{figure}
	\centering
	\includegraphics[width=\linewidth]{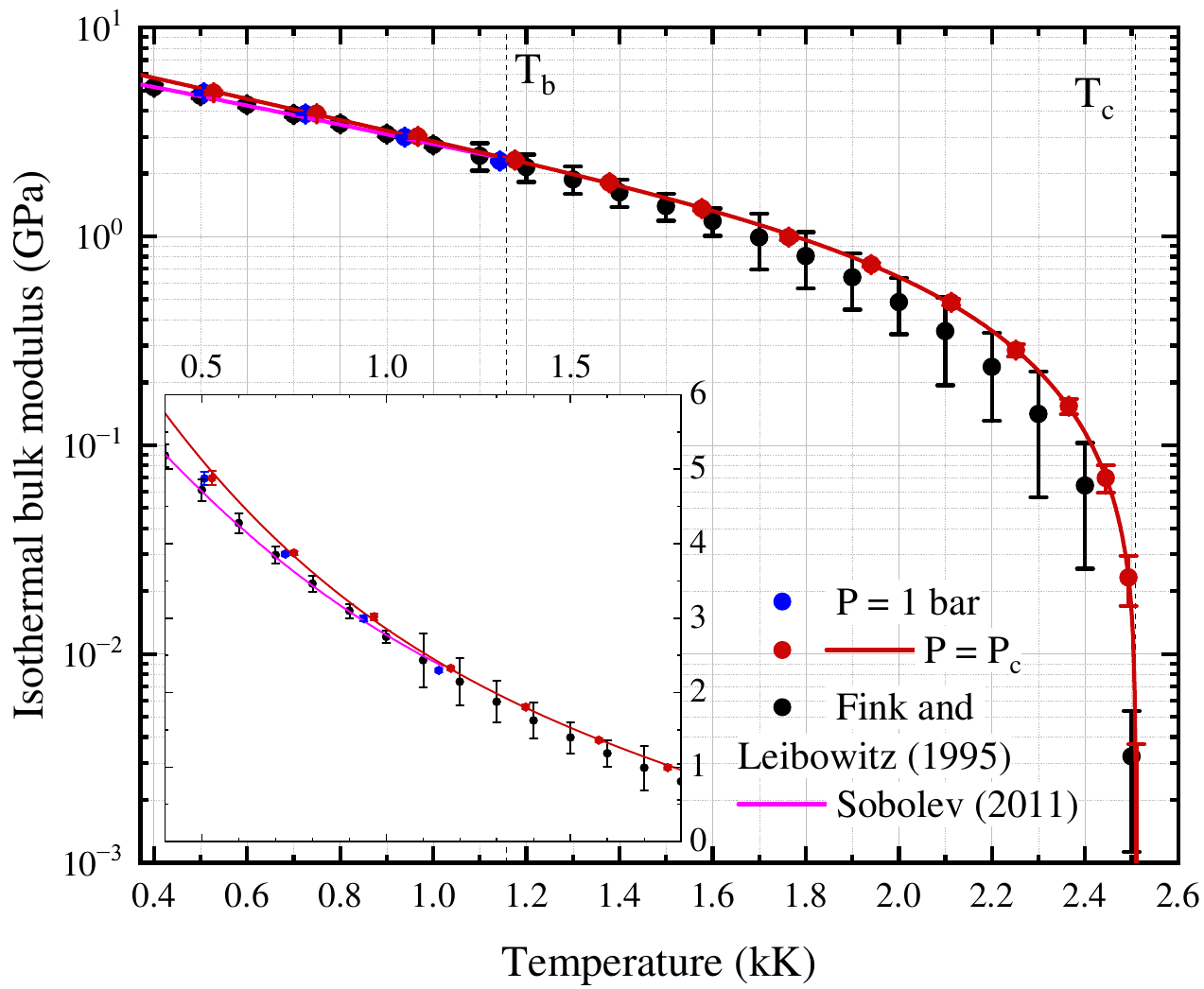}
	\caption{Isothermal bulk modulus $K_T$ of sodium along the atmospheric-pressure and critical isobars. Symbols show the DP-$r^2$SCAN results, and the solid curves are analytic approximations. Black symbols are values obtained from the Fink--Leibowitz saturated-liquid compressibility~\cite{FinkLeibowitz1995SodiumProperties}, and the magenta curve is the Sobolev normal-pressure dependence~\cite{Sobolev2011GENIVCoolants}.  The inset shows the low-temperature region on a linear ordinate scale.}
	\label{fig:Kt}
\end{figure}

The corresponding isentropic and isothermal bulk moduli are shown in Figs.~\ref{fig:Ks} and~\ref{fig:Kt}. Both quantities decrease continuously with temperature, indicating mechanical softening of the liquid. At low temperatures, the atmospheric-pressure and critical-isobar results closely follow the Sobolev normal-pressure dependence. Along the atmospheric-pressure isobar, $K_S$ and $K_T$ decrease from 5.59 and 4.87~GPa at 0.507~kK to 3.23 and 2.30~GPa at $T_b$, respectively. Along the critical isobar, the corresponding values decrease from 5.61 and 4.88~GPa at approximately 0.53~kK to 0.137 and 0.023~GPa at 2.493~kK, respectively. The Fink--Leibowitz values shown in Figs.~\ref{fig:Ks} and~\ref{fig:Kt} are obtained from the recommended saturated-liquid compressibilities. They are consistent with the present results at low temperatures, whereas their uncertainties increase strongly in the high-temperature region.

%The analytic critical-isobar approximations satisfy $K_S\rightarrow0$ and $K_T\rightarrow0$ as $T\rightarrow T_c$, with the stronger decrease of $K_T$ reflecting the divergence of the isothermal compressibility.

\begin{figure}
	\centering
	\includegraphics[width=\columnwidth]{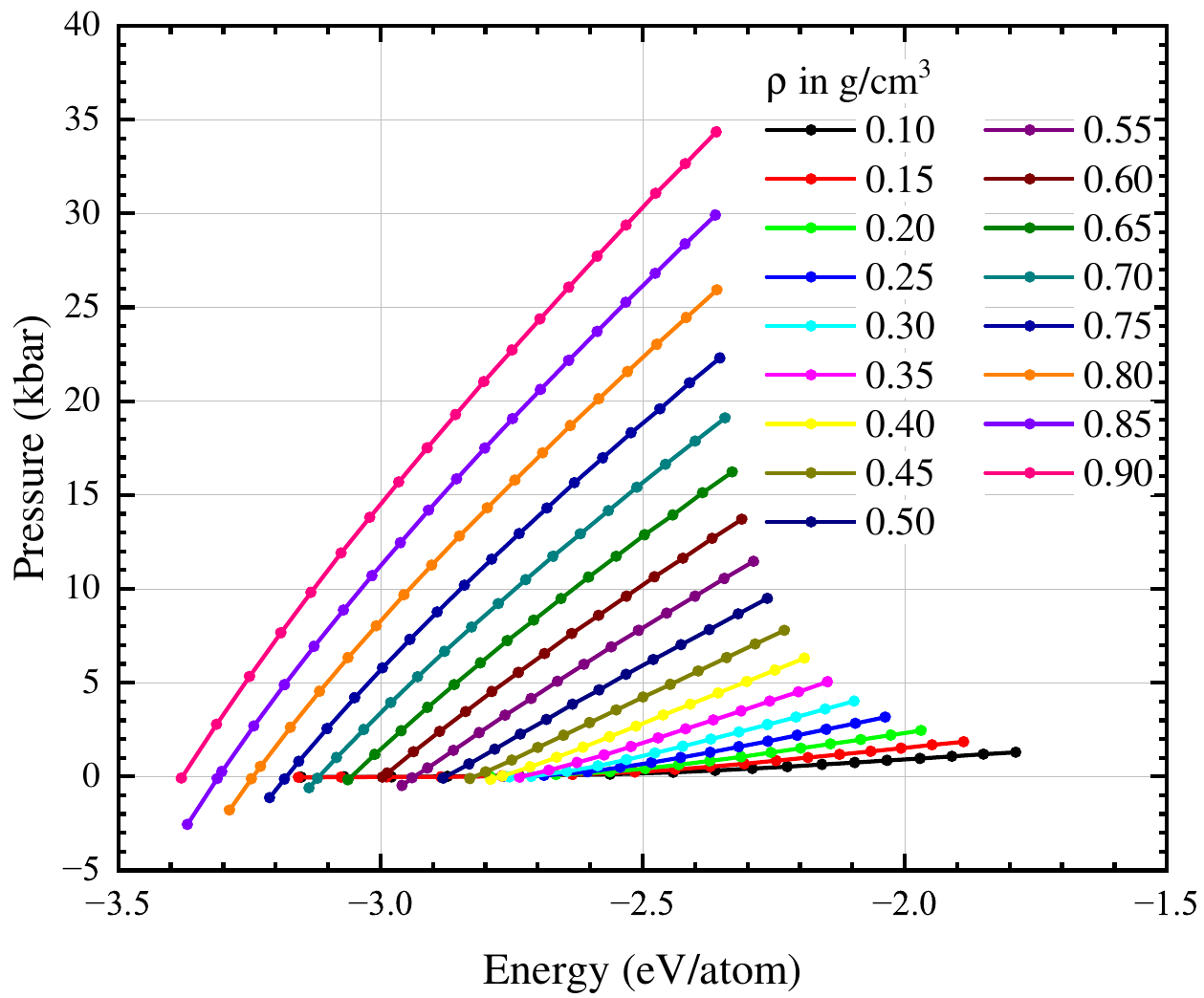}
	\caption{Pressure--energy isochores calculated with DP-$r^2$SCAN. Legend entries give the density in $\mathrm{g/cm^{3}}$. }
	\label{fig:isochores-pe}
\end{figure}

\begin{figure}
	\centering
	\includegraphics[width=\columnwidth]{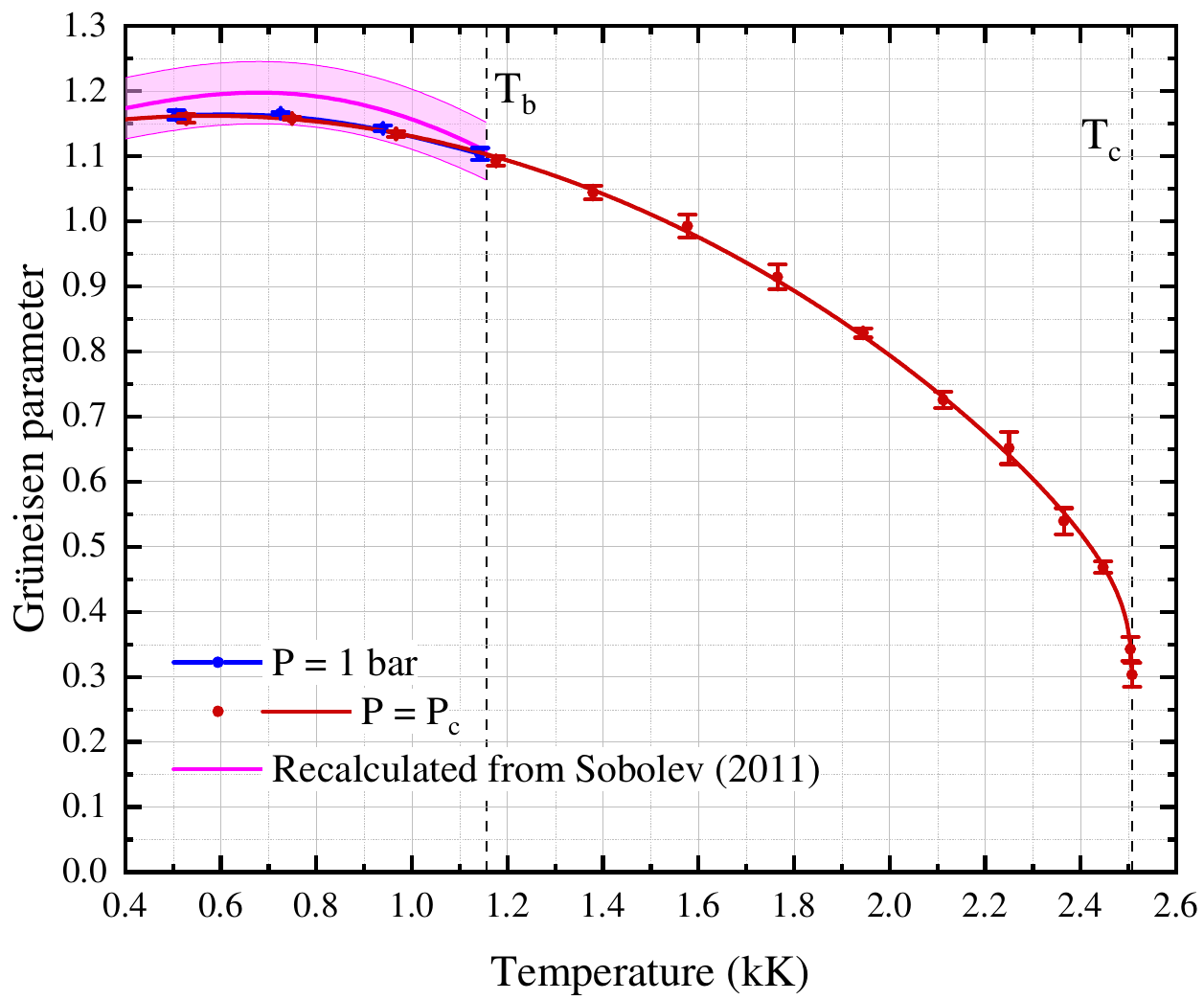}
	\caption{Gr\"uneisen parameter of sodium along the ambient-pressure and critical isobars. Symbols are calculated from the pressure--energy isochores, and solid lines connect the calculated values. The magenta curve is recalculated from the Sobolev normal-pressure thermophysical correlations~\cite{Sobolev2011GENIVCoolants}; the shaded region indicates its estimated uncertainty.}
	\label{fig:gamma-isobars}
\end{figure}

The Gr\"uneisen parameter is calculated directly from the pressure--energy dependence along the isochores (see Eq.~\eqref{eq:gruneisen}, Figs.~\ref{fig:isochores-pe} and~\ref{fig:gamma-isobars}).
The $P(E)$ curves at $V=\textrm{const}$ are shown in Fig.~\ref{fig:isochores-pe}. 

As shown in Fig.~\ref{fig:gamma-isobars}, the Gr\"uneisen parameter decreases monotonically with temperature. The normal-pressure and critical-isobar results remain very close throughout the equilibrium normal-pressure liquid range up to $T_{\mathrm b}$. Along the normal-pressure isobar, $\gamma$ decreases from approximately 1.17 at 0.507~kK to 1.103 at $T_b$. Along the critical isobar, it decreases from approximately 1.16 at 0.528~kK to about 0.3 in the vicinity of $T_c$.
For comparison, Fig.~\ref{fig:gamma-isobars} also shows $\gamma$ recalculated from the Sobolev normal-pressure thermophysical correlations using the same thermodynamic relation between $\gamma$, $\alpha_P$, $C_P$, and the sound velocity. This reconstructed dependence agrees with the present results in the low-temperature range.

\begin{figure}
	\centering
	\includegraphics[width=\columnwidth]{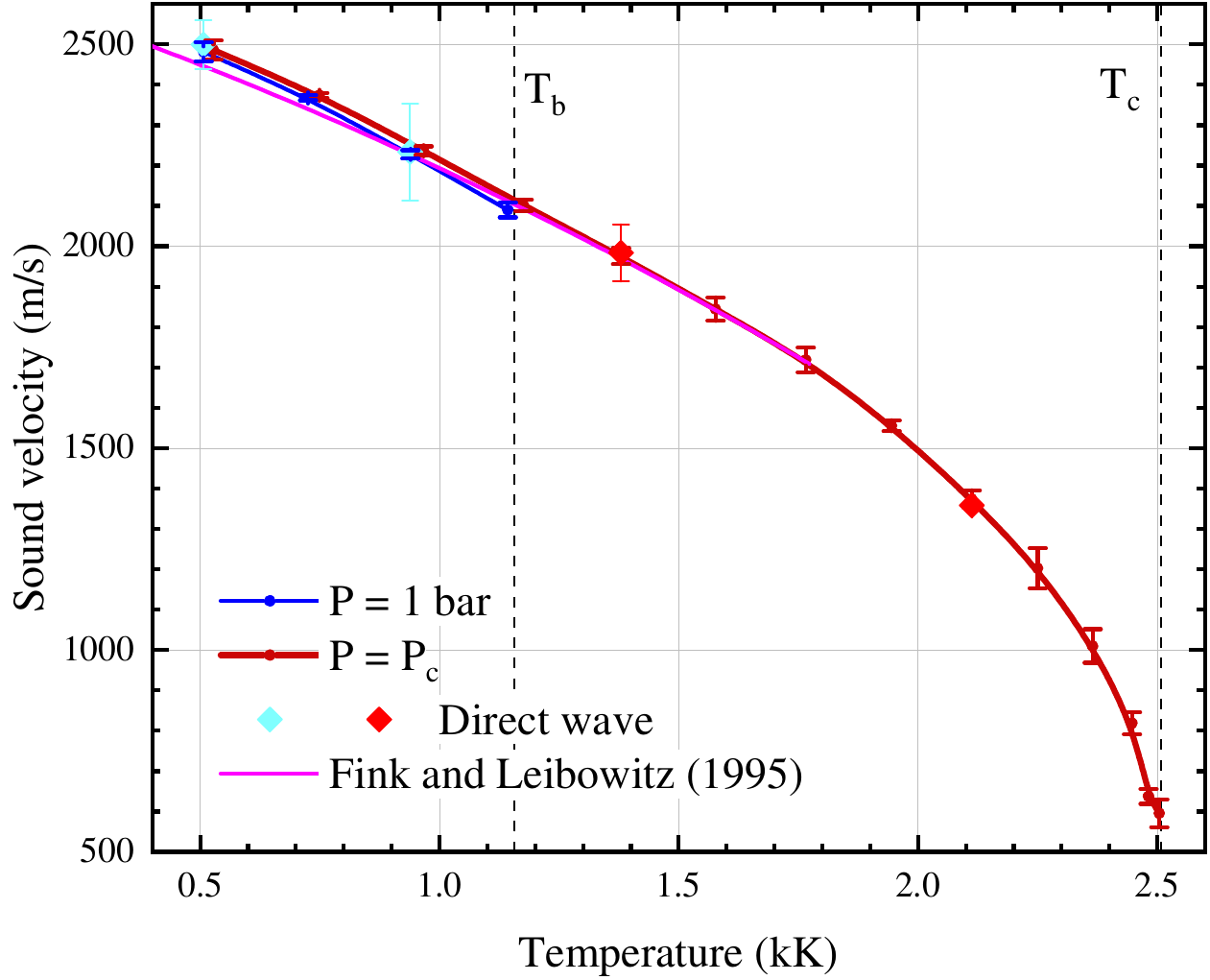}
	  \caption{Adiabatic speed of sound along the normal-pressure and critical isobars. Blue and red circles denote values calculated from the thermodynamic functions along the normal-pressure and critical isobars, respectively, and filled diamonds denote direct acoustic simulations. The magenta curve is the Fink--Leibowitz recommended dependence ~\cite{FinkLeibowitz1995SodiumProperties}.}
	\label{fig:sound-isobars}
\end{figure}

Using the density, $C_P$, $\alpha_P$, and $\gamma$ dependences obtained above, the adiabatic speed of sound is calculated from Eq.~\eqref{eq:sound-thermodynamic}. Figure~\ref{fig:sound-isobars} shows that the sound velocity decreases with temperature, following the progressive mechanical softening seen in the bulk moduli. Along the atmospheric-pressure isobar, the sound velocity decreases from approximately 2.5~km/s at 0.5~kK to 2.1~km/s at $T_b$. Along the critical isobar, the sound velocity continues to decrease and reaches approximately 0.60~km/s near the critical point.

The direct acoustic calculations provide an independent validation of the thermodynamic reconstruction. At 0.5 and 0.94~kK, the direct calculations give $2500(60)$ and $2200(100)$~m/s, respectively, in agreement with the thermodynamic values within their uncertainties. At lower temperatures, the calculated dependence also closely reproduces the Fink--Leibowitz recommendation~\cite{FinkLeibowitz1995SodiumProperties}. %The available measurements of Shaw and Caldwell were performed at different pressure--temperature conditions and are therefore not included in the direct comparison~\cite{ShawCaldwell1985SoundAlkali}.

\subsection{Phase boundaries and surface tension}\label{sec:phase-results}

%\subsubsection{Melting}\label{sec:normal-density-melting}

The melting temperature is determined using the two-phase $NVT$ simulations described in Sec.~\ref{sec:phase-methods}. The calculations cover the density range $\rho=0.93$--$0.98$~g/cm$^{3}$. For each density, the interface velocity changes sign within the selected temperature interval. Linear interpolation to $v_{\mathrm{int}}=0$ yields six points on the melting curve in the pressure range from $-1.65$ to $2.09$~kbar. These points are approximated by
\begin{equation}
P_{\mathrm m}(T_{\mathrm m})=M_1T_{\mathrm m}+M_0.
\label{eq:melting-line}
\end{equation}
Here $T_{\mathrm m}$ is in K and $P_{\mathrm m}$ is in kbar. The coefficients are $M_1=0.099$~kbar/K and $M_0=-34.2$~kbar. Interpolation of this dependence to atmospheric pressure with the bootstrap procedure gives
\begin{equation}
T_{\mathrm m,0}=346(2)\ \mathrm{K}.
\end{equation}
The slope of the approximation is in agreement with the experimental value 0.11~kbar/K~\cite{Gurvich1978ThermodynamicProperties,Boehler1983}. The calculated normal-pressure melting temperature is 25~K, or 6.7\%, below the recommended value $370.93(6)$~K~\cite{narayana2024alkali}. 
Menon \textit{et al.}~\cite{Menon2024Pyiron} show that a free-energy difference of 1~meV/atom may shift a transition temperature by up to about 50~K. In Ref.~\cite{Zhu2024TaVCrW}, electronic free-energy corrections of 4.2--5.8~meV/atom change the melting temperature by 48--57~K. The free-energy RMSE of the  DP-$r^2$SCAN model is 3.5~meV/atom. Thus, within the accuracy of the model, the calculated melting temperature agrees well with experiment.

%\subsubsection{Liquid--vapor coexistence and surface tension}\label{sec:interface}

\begin{figure}
	\centering
	\includegraphics[width=\columnwidth]{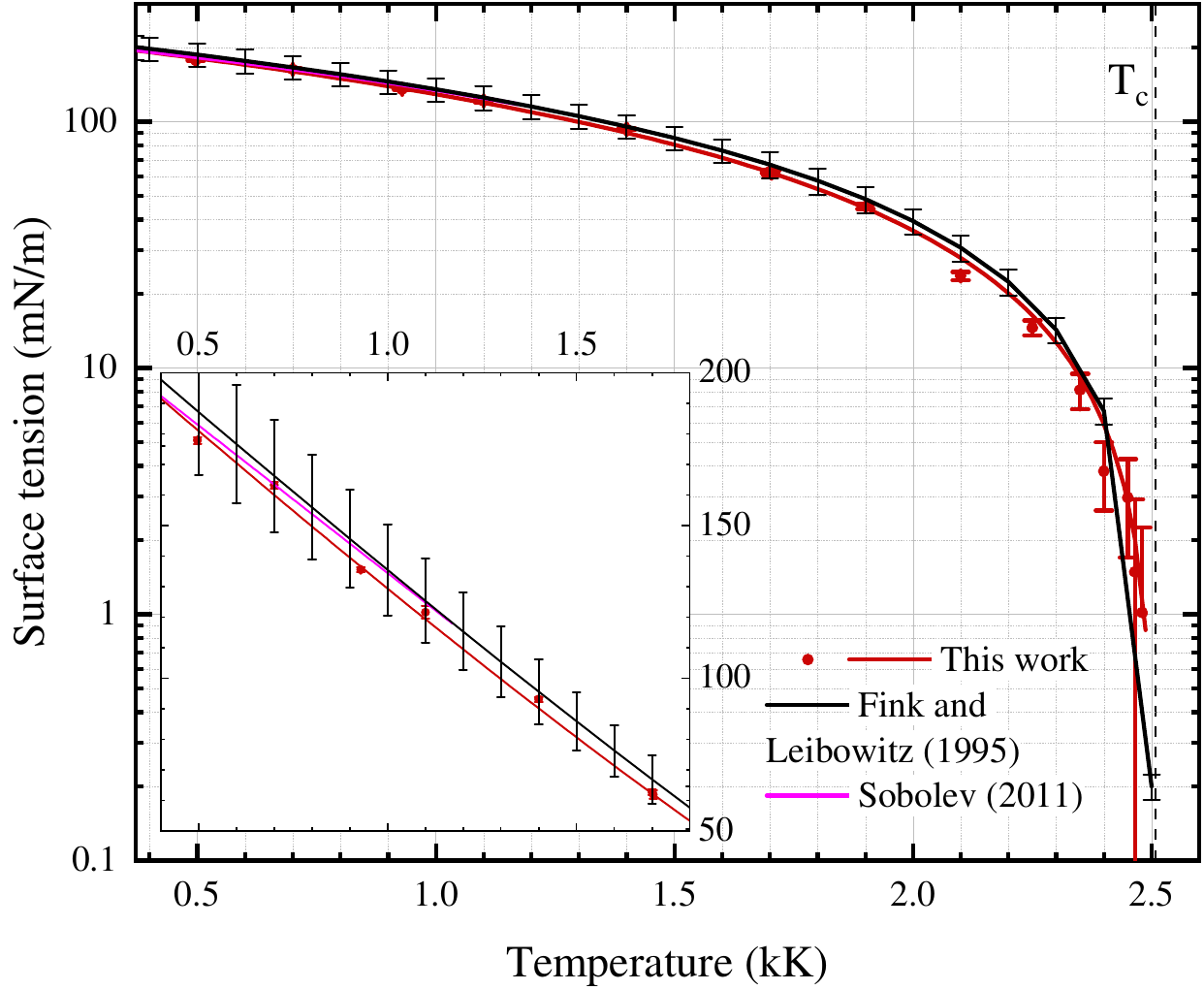}
	  \caption{Surface tension of sodium along the liquid--vapor coexistence curve. Red symbols denote the present DP-$r^2$SCAN results, and the red curve is the analytic approximation. The black and magenta curves show the Fink--Leibowitz and Sobolev recommended dependences, respectively~\cite{FinkLeibowitz1995SodiumProperties,Sobolev2011GENIVCoolants}. The inset shows the low-temperature region on a linear ordinate scale. }
	\label{fig:surface-tension}
\end{figure}

The coexistence densities obtained from the slab profiles using
Eq.~\eqref{eq:slab-density-profile} are shown in
Fig.~\ref{fig:nacritpointphasescan}. The same two-phase simulations
are used to calculate the surface tension from
Eq.~\eqref{eq:surface-tension-mechanical}; the resulting temperature
dependence is shown in Fig.~\ref{fig:surface-tension}. The calculated
surface tension decreases from 177.8~mN/m at 0.5~kK to
135.5~mN/m at 0.93~kK. At the same temperatures, the Sobolev
recommendation gives 183.0 and 141.3~mN/m, respectively
~\cite{Sobolev2011GENIVCoolants,Poindexter1929SodiumSurfaceTension,Kiriyanenko1965SodiumSurfaceTension}.
At higher temperatures, the surface tension decreases further to
61.9~mN/m at 1.703~kK and 1.02~mN/m at 2.480~kK, approaching
zero as the critical point is reached. Over the entire temperature
range, the calculated dependence closely follows the recommended
Fink--Leibowitz and Sobolev correlations and remains consistent
with their uncertainties.

\subsection{Transport properties of liquid sodium}\label{sec:dynamics}

\begin{figure}
\centering
\includegraphics[width=\columnwidth]{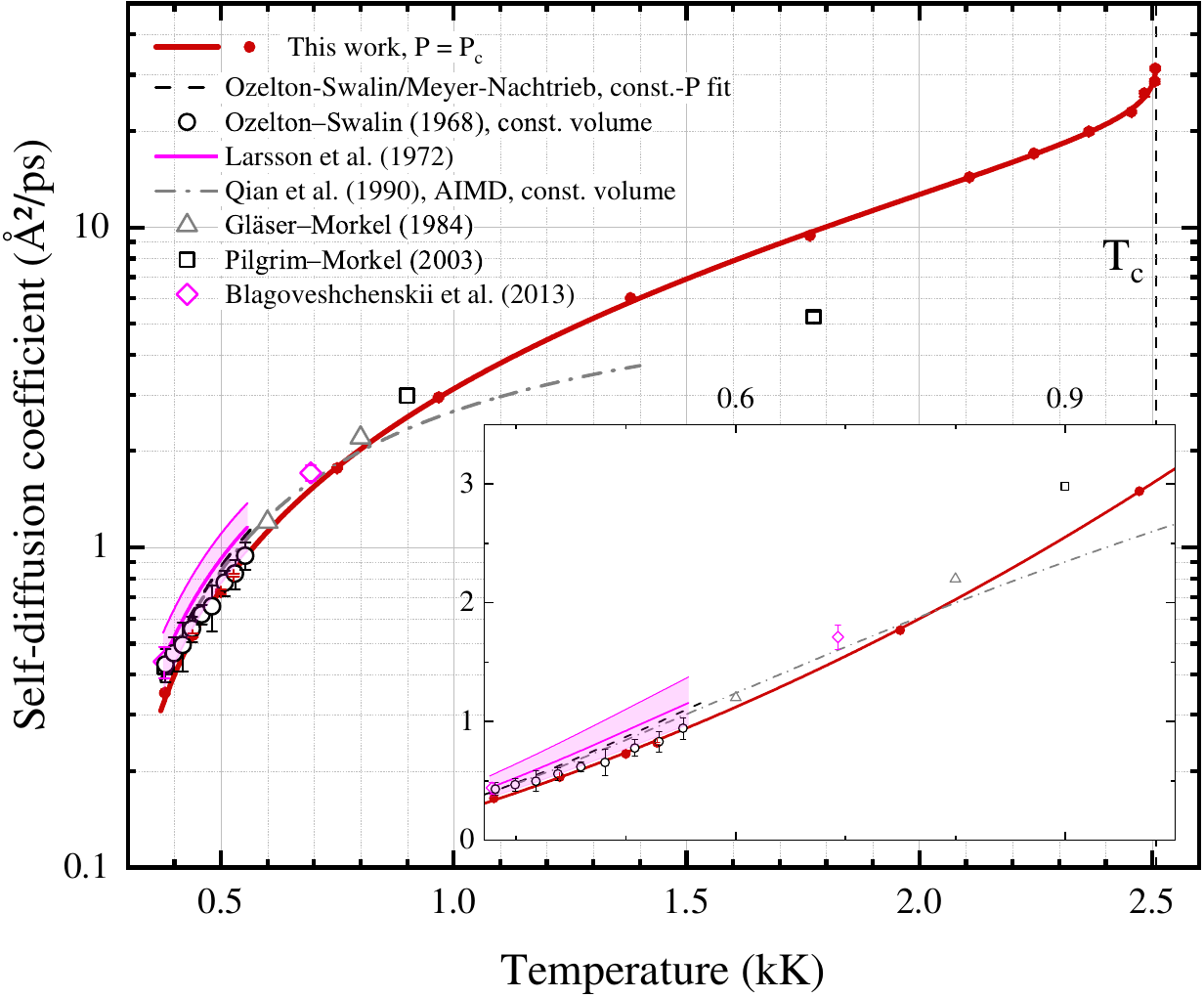}
\caption{Temperature dependence of the self-diffusion coefficient of liquid sodium.  Red stars with error bars denote the present DP-$r^2$SCAN results along the critical isobar, $P=P_c$, and the solid red line is the corresponding fit. The dashed black line is the constant-pressure Arrhenius correlation obtained by Ozelton and Swalin from their measurements combined with the earlier data of Meyer and Nachtrieb~\cite{MeyerNachtrieb1955SodiumDiffusion,OzeltonSwalin1968}; open circles show the constant-volume measurements of Ozelton and Swalin~\cite{OzeltonSwalin1968}. The magenta curve shows the correlation of Larsson \textit{et al.}~\cite{Larsson1972}. The gray dash-dotted curve denotes the \textit{ab initio} molecular-dynamics (AIMD) correlation of Qian \textit{et al.}~\cite{Qian1990NaAIMD}. Open gray triangles, open black squares, and open magenta diamonds indicate the literature data of Gl\"aser and Morkel~\cite{GlaeserMorkel1984}, Pilgrim and Morkel~\cite{PilgrimMorkel2003}, and Blagoveshchenskii \textit{et al.}~\cite{Blagoveshchenskii2014}, respectively. The constant-pressure Ozelton--Swalin curve and the Larsson curve are reconstructed from the published Arrhenius-type expressions, while the Qian line is reconstructed from the activation-energy relation given in their paper. The inset enlarges the low-temperature region.}
\label{fig:Na_self_diffusion}
\end{figure}

The simulation trajectories were analyzed to obtain the self-diffusion coefficients along the isochores, and the resulting data were then fitted to evaluate the diffusion coefficient along the critical isobar. Figure~\ref{fig:Na_self_diffusion} compares the present DP-$r^2$SCAN self-diffusion coefficients of liquid sodium along the critical isobar with available experimental and computational literature data. Along the critical isobar, the self-diffusion coefficient increases from 0.351~\AA$^2$/ps at 380~K to 0.726~\AA$^2$/ps at 500~K and then to about 30~\AA$^2$/ps in the critical region. At low temperatures, the present results agree well with the available experimental measurements. Somewhat higher values are given by the Meyer--Nachtrieb dependence~\cite{MeyerNachtrieb1955SodiumDiffusion}, which exceeds the two lowest-temperature DP-$r^2$SCAN values by approximately 20--24\%, and by the quasielastic-neutron result of Morkel and Pilgrim, 0.423~\AA$^2$/ps at 380~K~\cite{MorkelPilgrim2002SodiumDiffusion}.

For comparison, Fig.~\ref{fig:Na_self_diffusion} includes the constant-volume experimental data of Ozelton and Swalin~\cite{OzeltonSwalin1968} and their constant-pressure Arrhenius correlation obtained by combining their measurements with the earlier data of Meyer and Nachtrieb~\cite{MeyerNachtrieb1955SodiumDiffusion,OzeltonSwalin1968}, the tracer-diffusion correlation of Larsson \textit{et al.}~\cite{Larsson1972}, the quasielastic-neutron-scattering data of Gl\"aser and Morkel~\cite{GlaeserMorkel1984}, the high-temperature neutron-scattering results of Pilgrim and Morkel~\cite{PilgrimMorkel2003}, and the measurements of Blagoveshchenskii \textit{et al.}~\cite{Blagoveshchenskii2014}. In addition, the gray dash-dotted line shows the QMD correlation of Qian \textit{et al.}~\cite{Qian1990NaAIMD}. The Pilgrim--Morkel value at 1773~K is obtained by refitting the published quasielastic spectrum. Overall, the present DP-$r^2$SCAN results reproduce the available low-temperature experimental data well and follow the known experimental and first-principles trends over the broader temperature range, while extending the available diffusion data into the expanded-liquid and near-critical regions, where direct experimental information is sparse.

\begin{figure}
	\centering
	\includegraphics[width=\columnwidth]{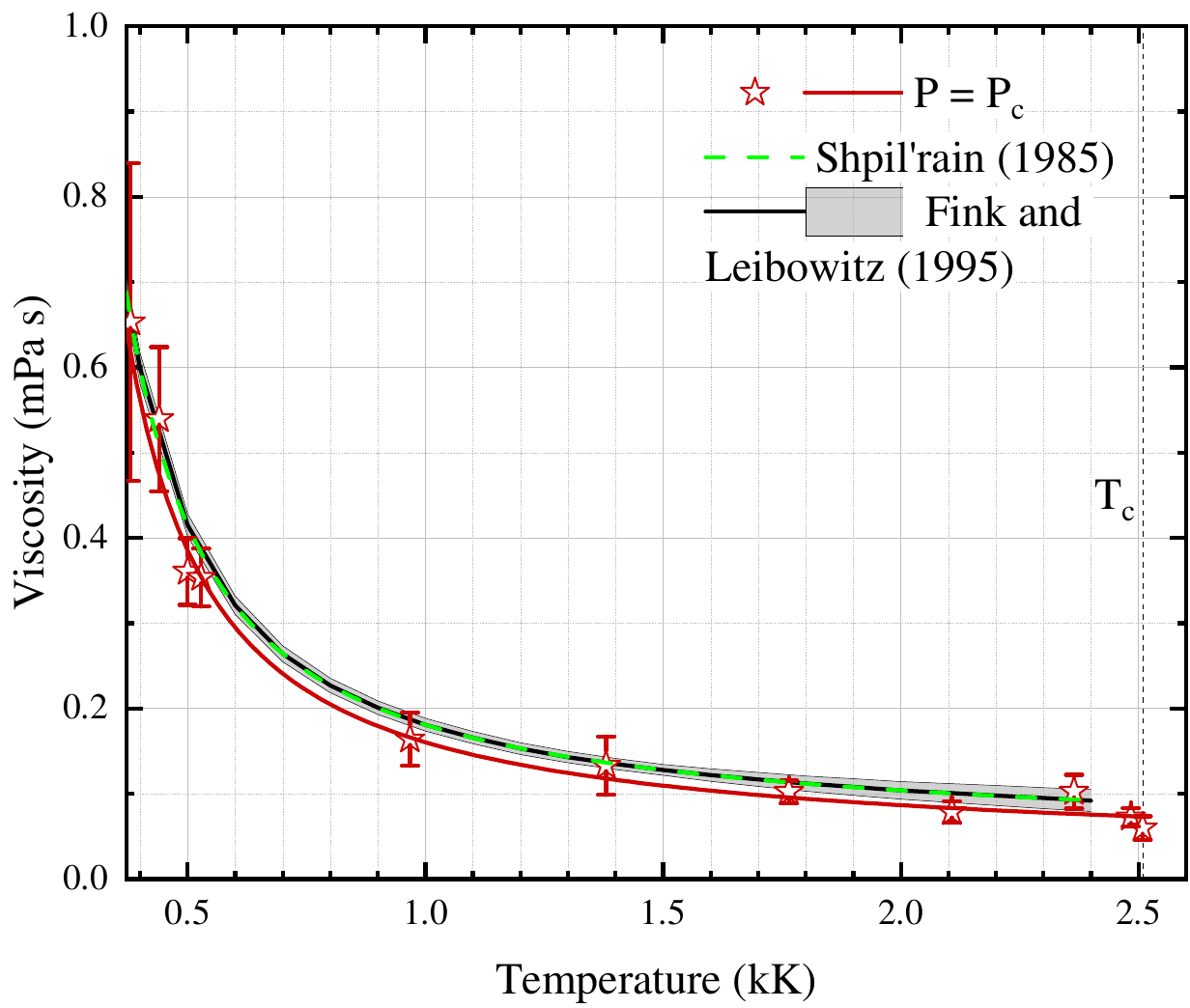}
  \caption{Shear viscosity of sodium along the critical isobar. Red stars show the DP-$r^2$SCAN results, and the solid red curve connects the calculated values. The green dashed curve is the Shpil'rain correlation~\cite{Shpilrain1985Viscosity}. The black curve and gray band show the Fink--Leibowitz recommended dependence and its uncertainty~\cite{FinkLeibowitz1995SodiumProperties}. The vertical dashed line marks the critical temperature $T_c$.}
	\label{fig:viscosity-critical}
\end{figure}

The shear viscosity calculated along the critical isobar is shown in
Fig.~\ref{fig:viscosity-critical}. It decreases from
0.653~mPa\,s at 0.38~kK to 0.06~mPa\,s at 2.5~kK.
The strongest decrease occurs in the low-temperature liquid region:
the viscosity falls rapidly below approximately 1~kK, whereas its
temperature dependence becomes considerably weaker at higher
temperatures. Near the critical point, the viscosity remains finite
and reaches values of about 0.06~mPa\,s.

The calculated values are compared with the Shpil'rain~\cite{Shpilrain1985Viscosity} 
and the Fink--Leibowitz~\cite{FinkLeibowitz1995SodiumProperties} recommended dependences. Over most of the
temperature range, the DP-$r^2$SCAN results closely follow both
literature dependences, although the calculated viscosity is
slightly lower. 
The literature correlations correspond to 
saturated-liquid conditions and are therefore used as reference
dependences rather than as a direct comparison with the critical
isobar.

\section{Conclusions}\label{sec:conclusions}

In this work, we construct temperature-dependent PBE, AM05, and $r^2$SCAN deep potentials for sodium from finite-temperature DFT data and use them to investigate the thermophysical properties from the vicinity of melting to the critical region. The comparison of the three exchange--correlation functionals shows a strong functional dependence of the predicted critical temperature. DP-PBE and DP-AM05 give $T_c=2.197(4)$ and 2.183(3)~kK, respectively, about 12\% below the recommended value, whereas DP-$r^2$SCAN gives $T_c=2.508(8)$~kK, $\rho_c=0.203(4)$~g/cm$^{3}$, and $P_c=0.249(6)$~kbar, in close agreement with the recommended critical parameters. The DP-$r^2$SCAN model also reproduces the normal bcc density within 0.06\% of the value obtained from the experimental lattice parameter. A separate electronic-entropy model is used to recover the internal energy required for the caloric properties.

Using the DP-$r^2$SCAN equation of state, we reconstruct the normal-pressure and critical isobars and obtain a consistent set of thermodynamic properties, including the enthalpy, $C_P$, $C_V$, thermal expansion coefficient, isothermal and isentropic bulk moduli, Gr\"uneisen parameter, and speed of sound. At low and intermediate temperatures, the calculated normal-pressure properties generally follow the available experimental data and recommended correlations. Along the critical isobar, on approaching the critical region, the thermal expansion and heat capacities increase, the bulk moduli decrease strongly, and the Gr\"uneisen parameter and speed of sound decrease. The speed of sound obtained from the thermodynamic functions is independently confirmed by direct acoustic simulations, providing a consistency check for the reconstructed EOS and its derivatives.

The phase-boundary calculations provide complementary tests of the potential. Direct solid--liquid coexistence simulations yield a local melting curve between $-1.65$ and 2.09~kbar with a slope of 0.099~kbar/K, close to the experimental value of 0.11~kbar/K. Its interpolation to ambient pressure gives $346(2)$~K, approximately 25~K below the recommended value. The magnitude of this deviation is consistent with the sensitivity of phase-transition temperatures to free-energy errors on the meV/atom scale. The calculated liquid--vapor coexistence densities reproduce the overall shape of the sodium binodal and approach the independently determined critical point. The surface tension agrees closely with the recommended dependences over the available temperature range, decreasing from 177.8~mN/m at 0.5~kK to 1.02~mN/m at 2.48~kK and approaching zero near $T_c$. 

The calculated transport properties extend the available information into the expanded-liquid and near-critical regions. Along the critical isobar, the self-diffusion coefficient increases strongly with temperature. The calculated values agree well with the available low-temperature experimental measurements and follow the known experimental and first-principles trends over the broader temperature range, while extending the available diffusion data into the near-critical region. The shear viscosity decreases from 0.653~mPa\,s at 0.38~kK to 0.06~mPa\,s near the critical point and remains close to the available reference dependences over most of the temperature range.

Thus, we demonstrate for the first time that a single DP-$r^2$SCAN neural-network interatomic potential can provide a unified description of sodium, simultaneously reproducing equilibrium thermodynamics, phase boundaries, interfacial and acoustic properties, and transport phenomena over an exceptionally broad thermodynamic range, from room temperature to the near-critical and supercritical regimes. The near-experimental agreement achieved across this diverse set of properties demonstrates an unprecedented level of transferability and predictive consistency for an atomistic model of sodium.

\begin{acknowledgments}
We thank Nikolay Chtchelkatchev for valuable discussions. The authors acknowledge the JIHT RAS Supercomputer Centre and the Shared Resource Centre ``Far Eastern Computing Resource'' IACP FEB RAS for providing computing time.
\end{acknowledgments}

%\bibliography{vaspref}


\end{document}